\documentclass[runningheads]{llncs}

\usepackage{graphicx}
\usepackage{comment}
\usepackage{amsmath,amssymb}
\usepackage{color}
\usepackage{url}
\usepackage{hyperref}

\usepackage[dvipsnames]{xcolor}
\usepackage{multirow}
\usepackage{mathtools}
\usepackage{tikz}
\usetikzlibrary{matrix}
\usepackage{adjustbox}
\usepackage{makecell}
\usepackage{float}
\usepackage[labelfont=bf,font=small,tableposition=bottom,labelsep=period]{caption}
\usepackage[skip=3pt]{subcaption}
\usepackage{booktabs}

\usepackage{tikz,pgfplots,pgf}

\pgfplotsset{width=10cm,compat=1.9}

\AtEndPreamble{
    \usepackage[capitalize]{cleveref}
    \crefname{section}{Sec.}{Secs.}
    \Crefname{section}{Section}{Sections}
    \crefname{table}{Tab.}{Tabs.}
    \Crefname{table}{Table}{Tables}
}

\usepackage{abbrv}

\definecolor{green(munsell)}{rgb}{0.0, 0.66, 0.47}
\definecolor{flamingopink}{rgb}{0.99, 0.56, 0.67}

\newif\ifreview
\reviewfalse

\ifreview
	\usepackage{lineno}

	\linenumbers
\fi

\usepackage{acronym}

\acrodef{vo}[VO]{visual odometry}
\acrodef{slam}[SLAM]{simultaneous localization and mapping}
\acrodef{gpu}[GPU]{graphic processor unit}
\acrodef{mla}[MLA]{micro lens array}
\acrodef{sfm}[SfM]{structure from motion}
\acrodef{rmse}[RMSE]{root mean square error}
\acrodef{sd}[SD]{standard deviation}

\begin{document}


\def\SubNumber{76}

\def\GCPRTrack{Fast Review Track}


\title{LiFCal: Online Light Field Camera Calibration via Bundle Adjustment}

\ifreview
\else

	\author{Aymeric Fleith\inst{1,2} \and
	Doaa Ahmed\inst{2} \and
	Daniel Cremers\inst{1} \and
	Niclas Zeller\inst{2}}
	
	\authorrunning{A. Fleith et al.}
	
	\institute{Technical University of Munich, Munich, Germany\\
    \email{\{aymeric.fleith, cremers\}@tum.de}\\
    \and Karlsruhe University of Applied Sciences, Karlsruhe, Germany\\
	\email{\{doaa.ahmed, niclas.zeller\}@h-ka.de}}
\fi

\maketitle              

\begin{abstract}
We propose LiFCal, a novel geometric online calibration pipeline for MLA-based light field cameras.
LiFCal accurately determines model parameters from a moving camera sequence without precise calibration targets, integrating arbitrary metric scaling constraints. It optimizes intrinsic parameters of the light field camera model, the 3D coordinates of a sparse set of scene points and camera poses in a single bundle adjustment defined directly on micro image points.

We show that LiFCal can reliably and repeatably calibrate a focused plenoptic camera using different input sequences, providing intrinsic camera parameters extremely close to state-of-the-art methods, while offering two main advantages: it can be applied in a target-free scene, and it is implemented online in a complete and continuous pipeline.

Furthermore, we demonstrate the quality of the obtained camera parameters in downstream tasks like depth estimation and SLAM.

Webpage: \url{https://lifcal.github.io/}.

\end{abstract}

\section{Introduction}
\label{sec:intro}

\begin{figure}[t]
  \centering
   \includegraphics[width=0.99\linewidth]{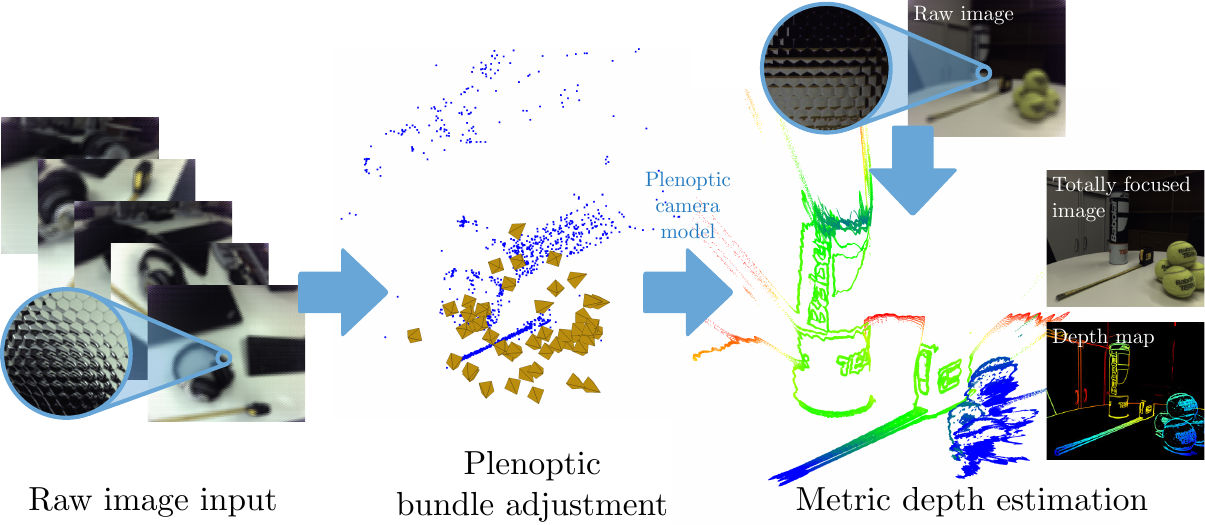}

   \caption{Process overview: Raw images undergo camera calibration via plenoptic bundle adjustment. This yields a metric camera model used to compute a totally focused image and depth map from new raw images, enabling accurate metric depth measurement.}
   \label{fig:processOverview}
\end{figure}

\Ac{vo} and \ac{slam} are expanding in fields like virtual reality, augmented reality, autonomous driving, using mainly monocular~\cite{engel2018direct,engel2014lsd}, stereo \cite{wang2017stereo} or RGB-D cameras~\cite{mur2017orb}.
Monocular cameras cannot determine scene scale without prior knowledge, stereo cameras are less stable in calibration, and RGB-D cameras need active scene illumination.

The concept of light field cameras has been around for many years~\cite{ives1903parallax,lippmann1908epreuves} and its value in depth estimation~\cite{wang2022occlusion}, super resolution~\cite{xiao2023cutmib}, \ac{vo} and \ac{slam}~\cite{zeller2018spo} is clear. What's more, it has a very large depth of field. Their appeal has grown with advancements in GPU processing and market availability, but cumbersome calibration has limited their mainstream adoption. As camera configurations change, target-free online calibration is essential for accurate data, \eg accurate 3D reconstruction. We address these shortcomings by introducing LiFCal, a new online calibration pipeline for \ac{mla}-based light field cameras, called plenoptic cameras in the sequel. To our knowledge, this is the first target-free calibration method for a complete plenoptic camera model, specifically for \ac{vo} or \ac{slam}. LiFCal offers the following key contributions:
\begin{itemize}
    \item An online calibration pipeline that accurately determines all intrinsic parameters of a plenoptic camera in scenes without a calibration target.
    \item A bundle adjustment formulation minimizing reprojection errors directly on micro image coordinates of a plenoptic camera and integrating arbitrary metric scaling constraints, acting as a complete \ac{sfm} pipeline for \ac{mla}-based plenoptic cameras.
    \item The integration of the plenoptic camera model into a micro-image-based depth estimation approach, providing undistorted metric depth maps and synthesized intensity images following the process described in \cref{fig:processOverview}.
\end{itemize}
We evaluate our pipeline's performance against state-of-the-art target-based plenoptic camera calibration on a public dataset~\cite{zeller2018dataset} and our own recordings. We demonstrate that our results closely match the reference calibration~\cite{zeller2018dissertation} with good precision and repeatability, enabling reliable metric point cloud generation.

The paper is organized as follows. \cref{sec:relatedWork} presents related work on plenoptic camera and online camera calibration. \cref{sec:method} introduces the LiFCal calibration method, covering the geometric model, the calibration pipeline, the initialization, and the plenoptic bundle adjustment. \cref{sec:evaluation} provides an extensive evaluation of the method and demonstrates the camera model's usability in tasks such as metric depth estimation and \ac{slam}. \cref{sec:conclusion} summarizes and concludes the work.

\section{Related Work}
\label{sec:relatedWork}

Our method enables online calibration of plenoptic cameras. This section covers both calibration for these cameras and online calibration for other cameras.

\subsection{Plenoptic Camera Calibration}
\label{sec:PlenopticCameraCalibration}

Several publications deal with the calibration of plenoptic cameras. They can be divided in two types: unfocused plenoptic cameras (or plenoptic camera 1.0) and focused plenoptic cameras (or plenoptic camera 2.0).

\subsubsection{Unfocused Plenoptic Cameras.} Unfocused plenoptic cameras were studied more closely in~\cite{adelson1992single,ng2005light}. The main lens is focused on the \ac{mla}, which is itself focused to infinity. The sensor plane is placed at the focal plane of the \ac{mla}.

A first method for correcting main lens aberrations on 4D light field data is introduced in~\cite{ng2006digital}. \cite{cho2013modeling} presents a full \ac{mla} calibration pipeline for Lytro cameras. The first complete mathematical model for general unfocused plenoptic cameras is introduced in~\cite{dansereau2013decoding}, including fifteen parameters and based on angle detection in sub-apertures images. It is refined by~\cite{zhou2019two} to represent physical camera parameters. A hybrid calibration approach by~\cite{darwish2019plenoptic}, calibrates sub-aperture images using geometric constraints of the main and micro lenses. \cite{ji2016light} uses a multi-view light field for calibration.

To improve feature detection in low-resolution micro images, \cite{bok2017geometric} uses line features from micro images, instead of the feature points used so far, and \cite{o2018calibrating,zhao2020metric} introduce a model using plenoptic discs.

\subsubsection{Focused Plenoptic Cameras.} For improved resolution, \cite{lumsdaine2009focused,lumsdaine2008full} explore the trade-off between spatial and angular information in light field data, leading to a focused plenoptic camera. The \ac{mla} is placed either in front or behind the image plane. Furthermore, the micro lenses are focused at the main lens's image plane rather than infinity, enhancing spatial resolution at the expense of angular precision.

The first calibration method for focused plenoptic cameras is introduced in~\cite{johannsen2013calibration}, using a fifteen-parameter model for a Raytrix camera and accounting for lateral image distortion. It is further refined in~\cite{heinze2016automated} with automatic calibration and an improved main lens distortion model, using a flat reference target.

In~\cite{zeller2014plencal}, three new models use the light field to relate object distance and virtual depth. Virtual depth calibration is added in~\cite{zeller2016depthcalib}. \cite{zeller2017plencalib} addresses lateral distortion after projecting virtual image points. The calibration approach~\cite{zeller2017dpo} estimates all intrinsic parameters, 3D object points, and camera poses in one task, reducing parameters to five but requiring a specific 3D target.

A calibration algorithm for multi-focus plenoptic cameras using only raw images is presented in~\cite{labussiere2020blur,labussiere2022leveraging} by introducing a new Blur Aware Plenoptic feature.

\cite{fachada2021calibration} calibrates the sub-aperture views considering images as a set of pinhole views. This approach is extended in~\cite{fachada2022pattern} for use without a reference pattern.

\subsection{Online Camera Calibration}
\label{sec:OnlineCameraCalibration}

Camera calibration typically requires a target. Online methods, however, determine intrinsic and extrinsic parameters without needing prior scene information.

Early target-free calibration methods~\cite{caprile1990using,pollefeys1997stratified} relied on scene assumptions and known environment structures.
Newer geometric approaches enable online calibration of monocular cameras in arbitrary scenes. In~\cite{schoenberger2016sfm}, the camera is calibrated while reconstructing the scene, determining parameters for each view.  \cite{lopez2019deep} uses a convolutional neural network to estimate camera parameters from a panorama. \cite{fang2022self} presents a self-supervised method using unconstrained image sequences.

Auto calibration extends to other camera types. 
\cite{zeisl2016structure} calibrates an RGB-D camera using sparse 3D reconstruction with \ac{sfm} or \ac{slam}. \cite{quenzel2017online} allows online depth calibration for an RGB-D camera using a visual \ac{slam} system without manual intervention. 
\cite{rehder2017online} introduces an online calibration system for stereo cameras, optimizing intrinsic and extrinsic parameters through bundle adjustment, starting with 3D scene reconstruction and trajectory estimation to optimize the parameters using a nonlinear solver. \cite{chang2024target} proposes a target-free calibration for a stereo camera-GNSS/IMU system, beginning with parameter initialization via reconstruction, followed by refinement with a solver.

\section{LiFCal Calibration Method}
\label{sec:method}

\begin{figure}[t]
  \centering
   \def\svgwidth{0.99\linewidth}
   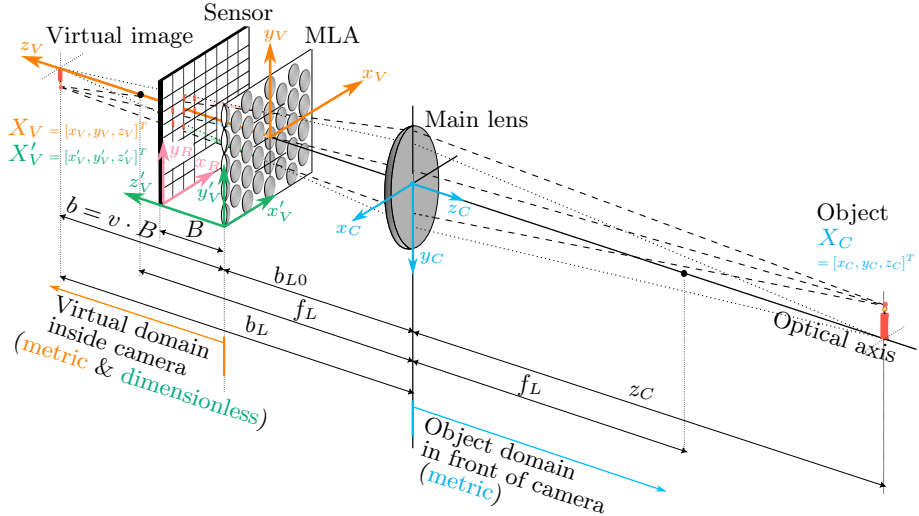

   \caption{Focused plenoptic camera in Galilean mode. Parameter $f_L$ is the main lens focal length, $b_{L0}$ the distance between main lens and \ac{mla}, $B$ the distance between \ac{mla} and sensor, $b$ the distance between \ac{mla} and virtual image, $b_L$ the distance between main lens and virtual image and $z_C$ the distance between real object and main lens. The virtual image created by the main lens is a mirror image of the real scene.}
   \label{fig:cameraModel}
\end{figure}

This section presents the plenoptic camera model (\cref{sec:plenopticCameraModel}) used for calibration and metric depth estimation, as background, inspired by~\cite{zeller2017dpo}, and the complete proposed calibration pipeline (\cref{sec:calibrationPipeline}), including initial calibration and plenoptic bundle adjustment.
The proposed LiFCal method does not require prior scene knowledge or targets, enabling better conditioning by using points with greater depth and spacing, overcoming limitations of simpler targets like checkerboards.

\subsection{Plenoptic Camera Model}
\label{sec:plenopticCameraModel}
Using a \ac{mla} between main lens and sensor (see \cref{fig:cameraModel}), a plenoptic camera captures a 4D light field in a single image.
The plenoptic camera model used in LiFCal is inspired by~\cite{zeller2017dpo} and employs a thin lens model for the main lens and a pinhole model for each micro lens. Lens distortion is defined directly on \ac{mla} and raw image coordinates.
While the camera model is introduced based on a focused plenoptic camera, it generally also holds for an unfocused plenoptic camera.
However, no manufacturers currently produce unfocused plenoptic cameras.

In the Galilean mode (see~\cite{lumsdaine2009focused} for the difference between Keplarian and Galilean modes), the main lens projects a 3D object space into a 3D virtual image space (see \cref{fig:cameraModel}).
A 3D point in object space is defined by its metric camera coordinates $X_C=[x_C, y_C, z_C]^T$ or homogeneous coordinates $\overline{X}_C=[x_C, y_C, z_C, 1]^T$.

Using the thin lens model, an object point $X_C$ can be projected into the virtual image space with coordinates $X_V=[x_V, y_V, z_V]^T$ (or homogeneous coordinates $\overline{X}_V=[x_V, y_V, z_V, 1]^T$).
The virtual image space, with its origin at the intersection of the main lens' optical axis and the \ac{mla}, is a mirrored coordinate system, meaning that its axes are all in the opposite direction compared to the camera frame.
Hence, the projection from $X_C$ to $X_V$ is defined as follows:
\begin{align}
	\begin{split}
		\lambda \cdot \overline{X}_V = K \cdot \overline{X}_c \qquad \Leftrightarrow \qquad
		\lambda \cdot  \begin{bmatrix}
			x_V\\
			y_V\\
			z_V\\
			1
		\end{bmatrix} = \begin{bmatrix}
			b_L & 0 & 0 & 0\\
			0 & b_L & 0 & 0\\
			0 & 0 & b & 0\\
			0 & 0 & 1 & 0
		\end{bmatrix} \cdot \begin{bmatrix}
			x_C\\
			y_C\\
			z_C\\
			1
		\end{bmatrix}
	\end{split}
	\label{eq:projectionXcToXv}.
\end{align}
The parameters $b_L$ and $b$ depend on the object point distance $z_C$.
The relation between $b_L$, $b$ and $z_C$ is expressed based on the thin lens equation in \cref{eq:bL}.

\begin{align}
	b_L=\left(\frac{1}{f_L}-\frac{1}{z_C}\right)^{-1}=b+b_{L0}
	\label{eq:bL}
\end{align}

A point in metric coordinates $X_V$ in the virtual image is converted to dimensionless coordinates $X_V'=[x_V', y_V', z_V']^T$.
$x_V'$ and $y_V'$ are defined in pixels (\cref{eq:projectionXvToXvPrime}), while $z_V'$ is defined by the so called virtual depth $v$ (\cref{eq:virtualDepth}), introduced in~\cite{perwass2012single} to generate depth maps from plenoptic images without metric calibration \cite{zeller2015plenopticdepth}. In \cref{eq:projectionXvToXvPrime}, $s_x$ and $s_y$ represent pixel size and $C_L=[c_x,c_y]^T$ is the principal point of the main lens.
While pixel dimensions are not essential, they provide a metric reference for the parameters.
\begin{align}
	x_V' &= x_V \cdot s_x^{-1}+c_x, \qquad y_V' = y_V \cdot s_y^{-1}+c_y
	\label{eq:projectionXvToXvPrime}\\
	v &=\frac{b}{B}=\frac{b_L-b_{L0}}{B} \quad \Rightarrow \quad b_L = v\cdot B+b_{L0}
	\label{eq:virtualDepth}
\end{align}

\begin{figure}[tb]
  \centering
  \begin{subfigure}{0.48\linewidth}
       \def\svgwidth{0.9\linewidth}
       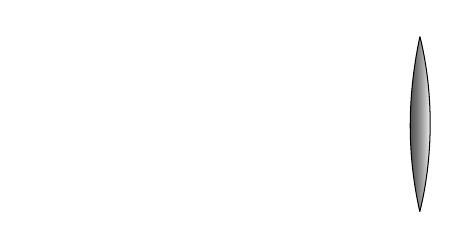
    
       \caption{Projection to a virtual image.}
       \label{fig:projectionDepth}
  \end{subfigure}
  \hfill
  \begin{subfigure}{0.48\linewidth}
       \def\svgwidth{0.9\linewidth}
       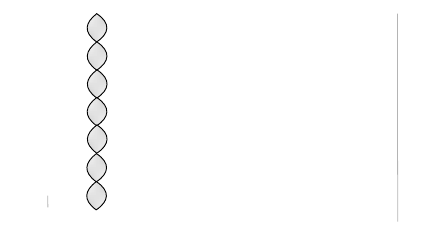
    
       \caption{Projection to the micro image.}
       \label{fig:centersProj}
  \end{subfigure}
  \caption{Projection modeling inside the plenoptic camera. (\protect\subref{fig:projectionDepth}) Projection of a virtual image point $X_V'=[x_V', y_V', z_V' = v]^T$ to a raw image point $X_R=[x_R, y_R]^T$ through the micro lens center $C_{ML}=[c_{MLx}, c_{MLy}]^T$. The virtual image point is given in dimensionless coordinates, \ie image distance $b=v \cdot B$ is normalized by the distance $B$.
  (\protect\subref{fig:centersProj}) Projection from the micro lens centers $C_{ML}$ to the micro image centers $C_I$.}
  \label{fig:short}
\end{figure}

A point $X_V'$ in the virtual image can be projected to multiple points $X_R=[x_R, y_R]^T$ in the raw image as defined in \cref{eq:projectionXV} (see \cref{fig:projectionDepth} for reference), \ie observations of the same virtual image point in multiple micro images. Here, $C_{ML}=[ c_{MLx}, c_{MLy}]^T$ is the center of the associated micro lens.
\begin{align}
  x_R = (x_V' - c_{MLx}) \cdot v^{-1} + c_{MLx}, \qquad y_R = (y_V' - c_{MLy}) \cdot v^{-1} + c_{MLy}
  \label{eq:projectionXV}
\end{align}

In reality, micro image centers $C_I=[c_{Ix}, c_{Iy}]^T$ are not directly aligned with micro lens centers $C_{ML}$ (see \cref{fig:centersProj}).
In \cref{eq:projectionXV}, the micro lenses centers $C_{ML}$ are derived from micro image centers $C_I$ using \cref{eq:projIToML}. The centers $C_I$ can be determined beforehand using a white image recorded with the plenoptic camera.
\begin{align}
  C_{ML} = 
\begin{bmatrix}
c_{MLx}\\c_{MLy}\\b_{L0}
\end{bmatrix} \coloneqq 
C_I \cdot \frac{b_{L0}}{b_{L0}+B} =
\begin{bmatrix}
c_{Ix}\\c_{Iy}\\b_{L0}+B
\end{bmatrix} \cdot \frac{b_{L0}}{b_{L0}+B}
  \label{eq:projIToML}
\end{align}

The presented camera model projects an object point $X_C$ to multiple points $X_R$ on the recorded raw image, \ie into multiple micro images.
To account for position imperfections due to lens distortion and sensor misalignment, a lens distortion model is added.
Unlike other plenoptic camera models, distortion is applied directly to the recorded raw image, \ie on the micro image coordinates and micro lens centers.
This way, the distortion model can be integrated directly into downstream tasks like image synthesis and depth estimation (see \cref{sec:downstreamTasks}).
Therefore, a complex model for virtual depth distortion \cite{heinze2016automated}, which in general cannot be inverted in closed form, is omitted.
The model employs radial symmetric and tangential distortion based on~\cite{Brown1966DecenteringDO} (see supplementary material for details), though it can be substituted with any other distortion model.

While the micro image centers can be obtained beforehand, in addition to the distortion parameters, the model leaves five unknown parameters to be determined and optimized for camera calibration: focal length $f_L$; main lens principal point $C_L=[c_x,c_y]^T$ in pixels; distance $B$ between the \ac{mla} and the sensor; distance $b_{L0}$ between the main lens and the \ac{mla}.

\subsection{Calibration Pipeline}
\label{sec:calibrationPipeline}

Unlike the eight-point algorithm, solving the two view geometry problem for a monocular camera, a closed-form solution cannot be built for the raw images of a plenoptic camera. Each pair of micro images provides a different fundamental matrix. Several point correspondences would be required per micro image pair, which is not possible because of their small size.

\begin{figure}[tb]
  \centering
   \includegraphics[width=0.99\linewidth]{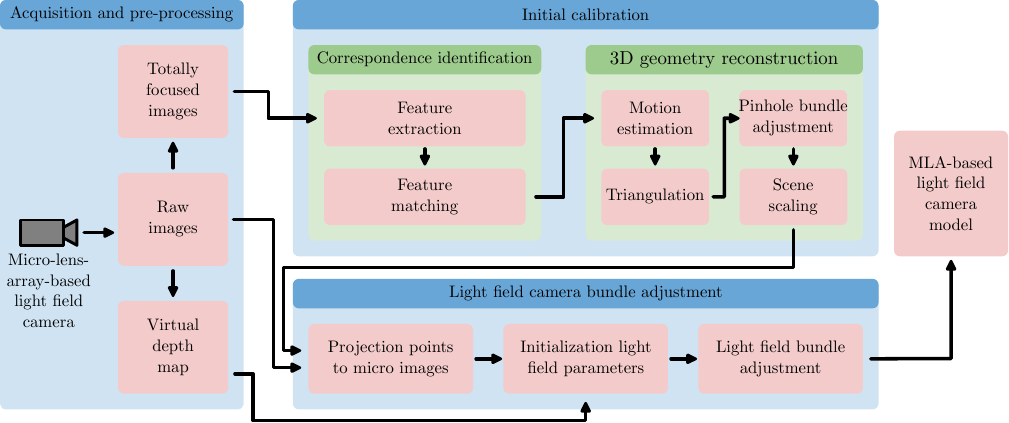}

   \caption{Flowchart of the online calibration algorithm for the focused plenoptic camera. Images are acquired from the camera and are first used to initialize the parameters. Next, a complete bundle adjustment for the plenoptic camera model is performed.}
   \label{fig:methodPipeline}
\end{figure}

The proposed pipeline, shown in \cref{fig:methodPipeline}, starts with acquiring raw data from the plenoptic camera (\cref{fig:processOverview}). From the estimated virtual depth $v$ and the raw image, we generate the so-called totally focused image and the virtual depth map~\cite{zeller2015plenopticdepth}. The totally focused image is only for feature detection to initiate bundle adjustment. Similarly, depth maps and raw images serve only for initialization, where precise values are not crucial, as parameters are optimized based on micro image coordinates $X_R$.
The initialization uses a pinhole camera model to get a first estimate of all camera poses, intrinsic parameters of the pinhole model and 3D point coordinates. The plenoptic model is then used to determine the exact intrinsic and extrinsic camera parameters and a sparse scene reconstruction in a plenoptic bundle adjustment formulation.

\subsubsection{Initial Calibration.}

For initialization, the plenoptic camera is modeled as a pinhole camera, estimating approximate values of all camera poses, intrinsic parameters of the pinhole model and 3D point coordinates~\cite{zeller2017dpo}.
SIFT features are extracted and matched from totally focused images, with outliers removed via RANSAC~\cite{fischler1981random}.
Assuming a static scene, initial camera poses $\xi \in \mathfrak{se}(3)$ are estimated for all views. 3D points are reconstructed through triangulation. Finally, the bundle adjustment jointly optimizes the 3D coordinates of the points $P \coloneqq \bigl\{X_W^{(1)}, \dots, X_W^{(N)} \bigr\}$, camera poses $\Xi \coloneqq \{\xi_1, \dots, \xi_M \}$ and intrinsic parameters of the pinhole model, remaining robust to noise. The first estimation is based on COLMAP~\cite{schoenberger2016sfm}, using a minimal version to reduce computation time.
During initialization, dimensions are defined at an arbitrary scale. Later, all data is scaled to metric using arbitrary scene constraints.

\subsubsection{Plenoptic Bundle Adjustment.}

Plenoptic bundle adjustment begins with initial pinhole model values for intrinsic and extrinsic parameters and 3D object points, which are then refined using Levenberg-Marquardt optimization.

The main lens focal length $f_L$ and principal point $C_L$ are set during initialization using the pinhole model, while $B$ and $b_{L0}$ are determined via a linear least squares problem based on the relationship between the virtual depth $v$ and the virtual image distance $b_L$ (see \cref{eq:virtualDepth}), set for every object point (see supplementary material for details).
Distortion parameters are initialized to zero. Optimal parameters for the model in \cref{sec:plenopticCameraModel} are then refined by minimizing the cost function in \cref{eq:costFunction} which can be adapted with any robust norm $\lVert \boldsymbol{\cdot} \rVert_{\gamma}$.
\begin{align}
  E(\Pi, \Xi, P) = \sum_{i=1}^{N} \sum_{j=1}^{M} \sum_{k=1}^{L} \lVert r_{(i,j,k)} \rVert_{\gamma} \cdot \theta_{(i,j,k)}
  \label{eq:costFunction}
\end{align}
In \cref{eq:costFunction}, $\Pi$ includes intrinsic and distortion parameters, $\Xi$ is the set of camera poses relative to a shared world frame ($\Xi \coloneqq \{\xi_1, \dots, \xi_M \}$), $P$ is the set of 3D reference points in the world frame ($P \coloneqq \bigl\{X_W^{(1)}, \dots, X_W^{(N)} \bigr\}$). The object point $X_W \in P$ in the world frame, \eg the frame of the first sequence view, can be converted to the camera frame of the $j$-th view by the rigid body transform $G(\xi_j) \in SE(3)$. The masking function $\theta_{(i,j,k)}$ is 1 if the $i$-th point of the object is visible in the $k$-th micro image of the $j$-th view, and is 0 otherwise. The residual vector $ r_{(i,j,k)}$ for a point with measured coordinates $X_{Rd}$, corresponding to the point of object $i$ seen in micro image $k$ of camera view $j$, is defined by \cref{eq:residuals}. The function $\pi_{ML}\left(\boldsymbol{\cdot}\right)$ projects camera coordinates to the micro image $k$ with the model defined in \cref{sec:plenopticCameraModel}. This nonlinear optimization problem is solved using the Levenberg-Marquardt algorithm, implemented by the Ceres Solver Library~\cite{Agarwal_Ceres_Solver_2022}.
\begin{align}
  r_{(i,j,k)} = \pi_{ML}\left(G(\xi_j)X_W^{(i)},C_{ML}^{(k)},\Pi\right) - X_{Rd}.
  \label{eq:residuals}
\end{align}

\section{Evaluation}
\label{sec:evaluation}

Several experiments were carried out to evaluate the performance of the proposed calibration pipeline, LiFCal, and to assess the quality of the resulting plenoptic camera model.
Furthermore, to demonstrate the high quality and usability of the obtained plenoptic camera model, the model is applied in downstream tasks like metric depth estimation and \ac{slam}. Here, we tightly integrate the plenoptic camera model into the depth estimation pipeline.

\subsection{Calibration}

Obtaining ground truth intrinsic parameters for any real camera model is almost impossible, and even more difficult for the complex model of a plenoptic camera.
Therefore, we compare the results of LiFCal against a state-of-the-art calibration pipeline~\cite{zeller2018dissertation} that is utilizing a complex 3D calibration target and professional photogrammetric software to obtain high quality camera parameters.
Several experiments were conducted to assess the repeatability and accuracy of the intrinsic camera parameters obtained by LiFCal.

\subsubsection{Calibration Based on a 3D Calibration Target.}

\setlength{\tabcolsep}{5pt}
\begin{table}[tb]
	\caption{Comparison of the intrinsic parameters of the plenoptic camera estimated by the reference method using a professional 3D calibration target and LiFCal. Deviations given in percentage are values relative to the respective reference calibration parameter.}
	\label{tab:parametersComparison}
	\centering
	\begin{tabular}{@{}l r | c c c c c}
		\toprule
		Method & Lens & $f_L$ [mm] & $b_{L0}$ [mm] & $B$ [mm] & $c_x$ [pixel] & $c_y$ [pixel] \\
		\midrule
		\multirow{3}{*}{\makecell[l]{Reference\\calibration\\(ground truth)}} & 12.5~mm & 13.181 & 12.209 & 0.399 & 1006.5 & 1041.5 \\
		& 16~mm & 16.748 & 15.893 & 0.376 & 1018.7 & 1054.2 \\
		& 35~mm & 35.368 & 34.471 & 0.370 & 1022.0 & 1031.0 \\
		\midrule
		\multirow{3}{*}{\makecell[l]{LiFCal\\(our method)}} & 12.5~mm & 13.193 & 12.228 & 0.395 & 1006.5 & 1041.6 \\
		& 16~mm & 16.745 & 15.893 & 0.375 & 1019.4 & 1053.5 \\
		& 35~mm & 35.342 & 34.445 & 0.369 & 1024.0 & 1028.5 \\
		\midrule
		\midrule
		\multirow{3}{*}{\makecell[l]{Deviation of\\LiFCal from the\\ reference [\%]}} & 12.5~mm & 0.091 & 0.155 & 1.003 & 0.000 & 0.010 \\
		& 16~mm & 0.018 & 0.000 & 0.266 & 0.069 & 0.066 \\
		& 35~mm & 0.074 & 0.075 & 0.270 & 0.196 & 0.242 \\
		\bottomrule
	\end{tabular}
\end{table}

In \cite{zeller2018dissertation,zeller2018dataset}, the plenoptic camera is calibrated using a one-meter cube 3D target together with a professional photogrammetric measurement software. LiFCal is applied to the same calibration images for comparison with this state-of-the-art method. Sample images can be found in the supplementary material.
\cite{zeller2018dissertation} provides calibration data with a Raytrix R5 camera (model: MG042CG-CM-TG) with three different main lenses: $f_L = 12.5$~mm, $f_L = 16$~mm, and $f_L = 35$~mm.
The camera has a resolution of 2048~pixels~$\times$~2048~pixels and a pixel size of 5.5~µm.
We use 94 images for the 12.5~mm lens, 70 images for the 16~mm lens, and 76 images for the 35~mm lens. The scene scale is obtained from known reference distances between markers on the target.

\cref{tab:parametersComparison} compares the intrinsic camera parameters obtained by LiFCal with the reference calibration \cite{zeller2018dissertation}. All parameters are estimated with relative errors below 0.3\%, except for parameter $B$ for the 12.5~mm lens (error of 1.003\%), which can be explained by the small focal length compared to the object distance. The parameters' accuracy is confirmed by the focal length $f_L$ closely aligning with nominal manufacturer values. Also, the fixed manufacturing parameter $B$ is consistent across estimates, with a small standard deviation of 0.013~mm. LiFCal achieves comparable results to~\cite{zeller2018dissertation} without using knowledge of the coded target markers, relying instead on standard SIFT features. While it struggles on homogeneous surfaces, it remains robust even with fewer features and images (see supplementary material for results with three different lenses).

\subsubsection{Online Calibration on Target-free Scenes.}

To demonstrate the online calibration performance of LiFCal, we use sequences from a plenoptic \ac{vo} dataset~\cite{zeller2018dataset} captured with a Raytrix R5 camera (same as for previous experiment) with a nominal focal length of $f_L=16$~mm.
We take images at regular intervals of the winding movement at the start of three sequences: Lab (seq\_004, using 56 images), Hallway (seq\_007, using 70 images), Office (seq\_009, using 76 images) (see supplementary material for samples).
The scene scale is obtained from a real distance measured on a visible object in the images and applied as a scale factor to the scene.

\setlength{\tabcolsep}{9pt}
\begin{table}[tb]
  \centering
  \caption{Estimated R5 camera parameters on target-free scenes and \ac{rmse} relative to the reference. The \ac{rmse} compares acquired data with reference from \cref{tab:parametersComparison}. It is expressed as a percentage error relative to the reference because of significant differences in orders of magnitude among parameters (\eg, nearly a factor of 50 between $B$ and $f_L$) to interpret the error with respect to the parameter.}
  \label{tab:calibrationAnyScene}
  \begin{tabular}{c | c c c c c}
    \toprule
    Scene & $f_L$ [mm] & $b_{L0}$ [mm] & $B$ [mm] & $c_x$ [pixel] & $c_y$ [pixel] \\
    \midrule
    Lab & 16.609 & 15.889 & 0.338 & 1022.7 & 1056.9\\
    Hallway & 16.788 & 15.889 & 0.384 & 1019.3 & 1050.0\\
    Office & 16.771 & 15.881 & 0.379 & 1021.9 & 1045.8 \\
    \bottomrule
    \ac{rmse} [\%] & 0.505 & 0.048 & 5.981 &0.292 & 0.535 \\
    \bottomrule
  \end{tabular}
\end{table}

The dataset contains sequences recorded only with the 16~mm lens, limiting evaluation in \cref{tab:calibrationAnyScene} to this configuration. However, the results can be compared with the 16~mm lines in \cref{tab:parametersComparison}.
The \ac{rmse} relative to the reference across all three calibrations is well below 0.6\% for all parameters, except $B$ as before.

\subsubsection{Online Recalibration.}

The plenoptic-camera-based online calibration still requires a reference scale in the scene to obtain metric model parameters.
However, online calibration also allows the readjustment of model parameters \eg obtained for a previous factory calibration.
While some parameters may change and become inaccurate, others, such as the main lens focal length $f_L$ and the distance $B$ between the \ac{mla} and the sensor, remain fixed and are sensor and lens specific respectively.
Therefore, in this experiment, we fix these parameters to reference values from~\cite{zeller2018dataset}, obviating additional steps for scene scale estimation. Pixel size becomes nonessential as parameters adapt accordingly, either in pixels or specified metric units. The sequences used are exactly the same as for the previous experiment.

\setlength{\tabcolsep}{9pt}
\begin{table}[tb]
  \centering
  \caption{Estimation of R5 camera parameters with recalibration by setting parameters $f_L = 16.748$~mm and $B=0.376$~mm, and comparison with the reference in \cref{tab:parametersComparison}.}
  \label{tab:onlineRecalibration}
  \begin{tabular}{c | c c c | c c c}
    \toprule
    \multirow{2}{*}{Scene} & \multicolumn{3}{c}{Estimated parameters} & \multicolumn{3}{ c }{Deviation} \\
     & \makecell{$b_{L0}$ $[$mm$]$} & \makecell{$c_x$ $[$pixel$]$} & \makecell{$c_y$ $[$pixel$]$} & \makecell{$b_{L0}$ $[$\%$]$} & \makecell{$c_x$ $[$\%$]$} & \makecell{$c_y$ $[$\%$]$} \\
    \midrule
    Lab & 15.944 & 1014.6 & 1057.8 & 0.32 & 0.40 & 0.34 \\
    Hallway & 15.869 & 1019.3 & 1051.6 & 0.15 & 0.06 & 0.25 \\
   Office & 15.866 & 1024.4 & 1053.4 & 0.17 & 0.56 & 0.08 \\
    \bottomrule
  \end{tabular}
\end{table}

\cref{tab:onlineRecalibration} compares the optimized variable parameters and their reference values, demonstrating errors below 0.6\% compared to the reference calibration. Here, no scene information is required, not even the scale.
Nevertheless, the results are extremely close to the other experiments and to the reference.

\subsubsection{Plenoptic Bundle Adjustment Accuracy.}

\begin{figure}[tb]
	\centering
	\centering
	\begin{subfigure}{0.3\linewidth}
		\includegraphics[width=1.0\linewidth]{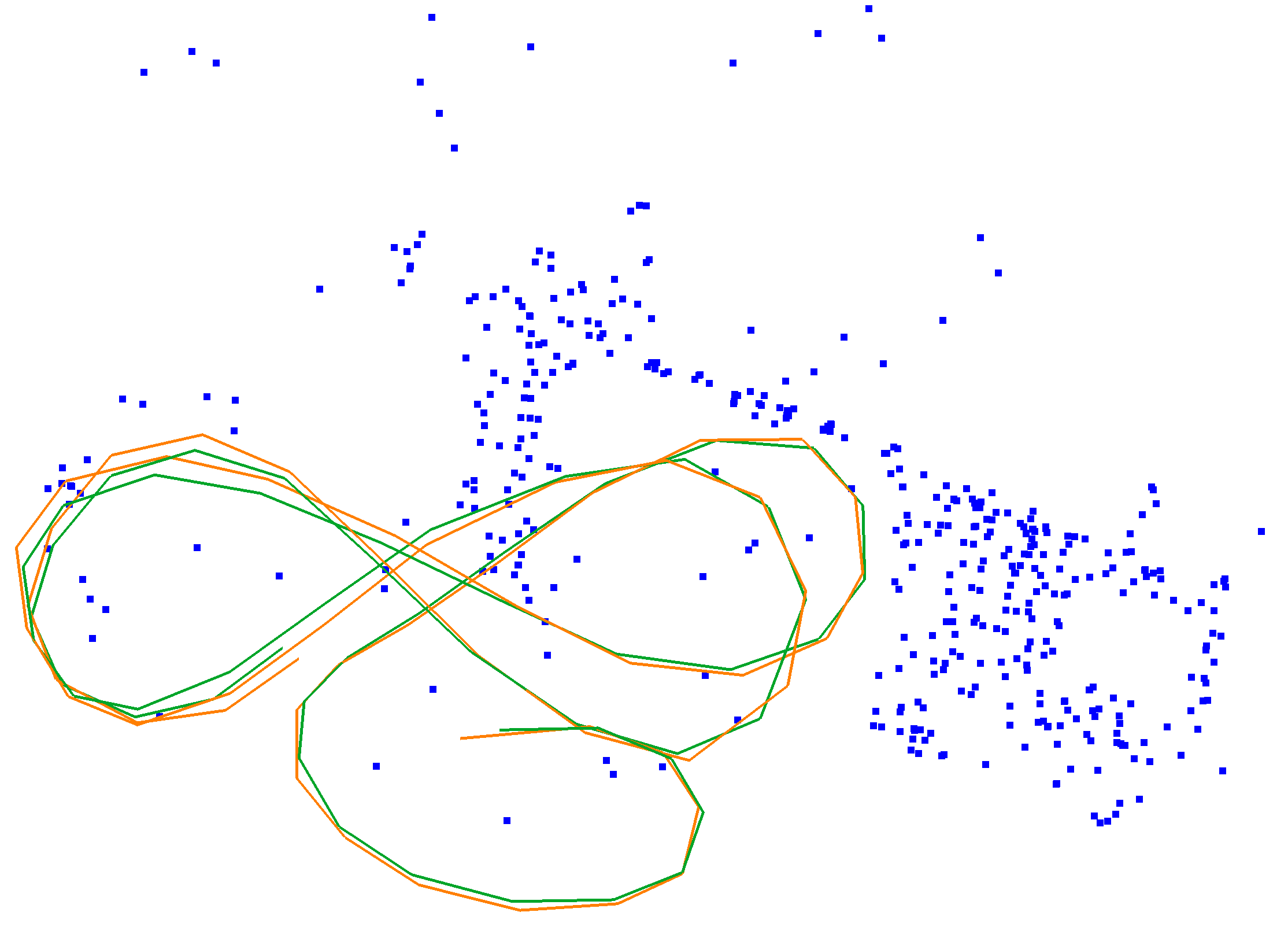}
		\captionsetup{justification=centering}
		\caption{Lab Sequence\\$\text{\ac{rmse}}=7.419~\text{mm}$}
		\label{fig:trajectories-a}
	\end{subfigure}
	\hfill
	\begin{subfigure}{0.3\linewidth}
		\includegraphics[width=1.0\linewidth]{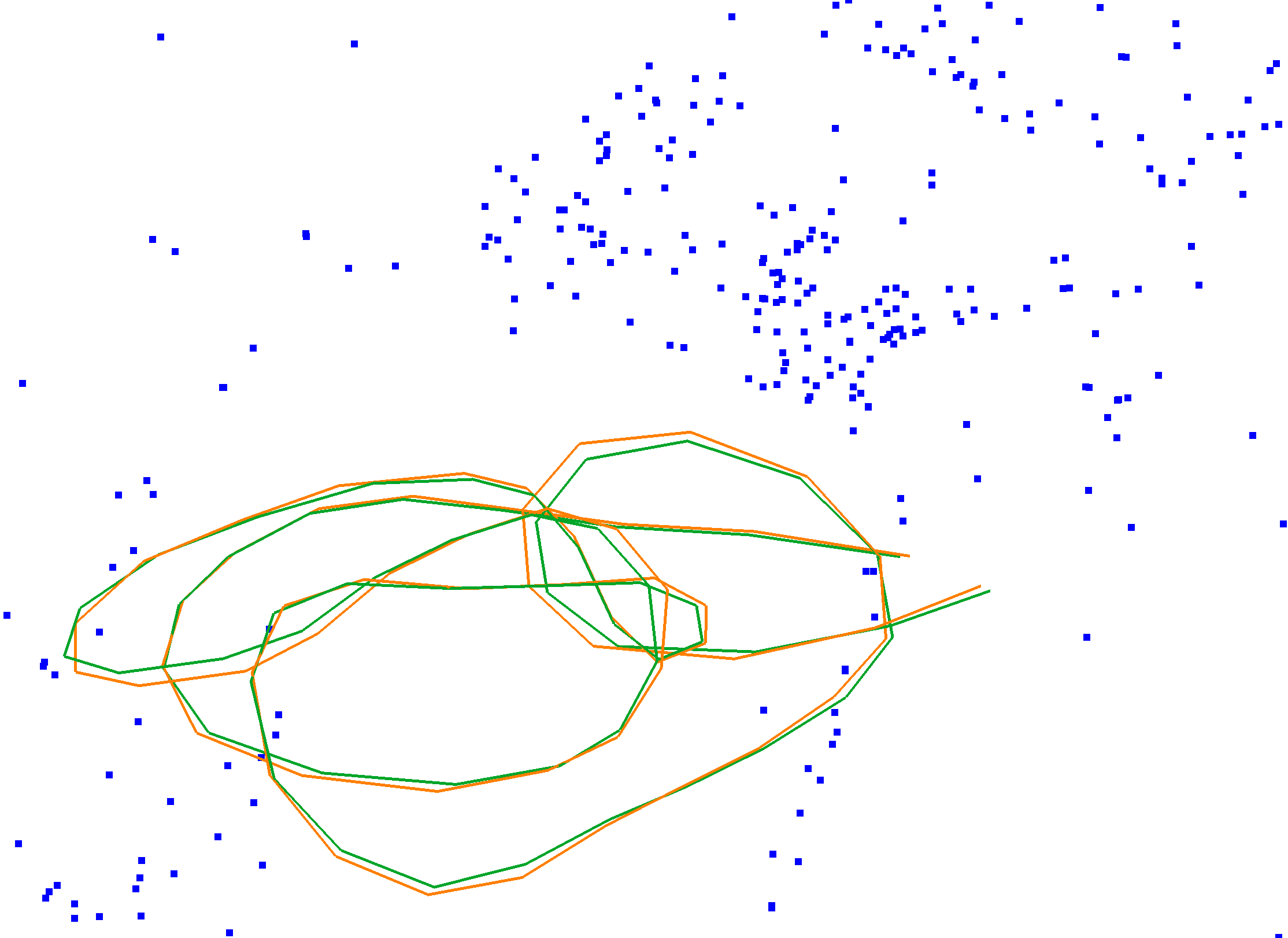}
		\captionsetup{justification=centering}
		\caption{Hallway sequence\\$\text{\ac{rmse}}=12.611~\text{mm}$}
		\label{fig:trajectories-b}
	\end{subfigure}
	\hfill
	\begin{subfigure}{0.3\linewidth}
		\includegraphics[width=1.0\linewidth]{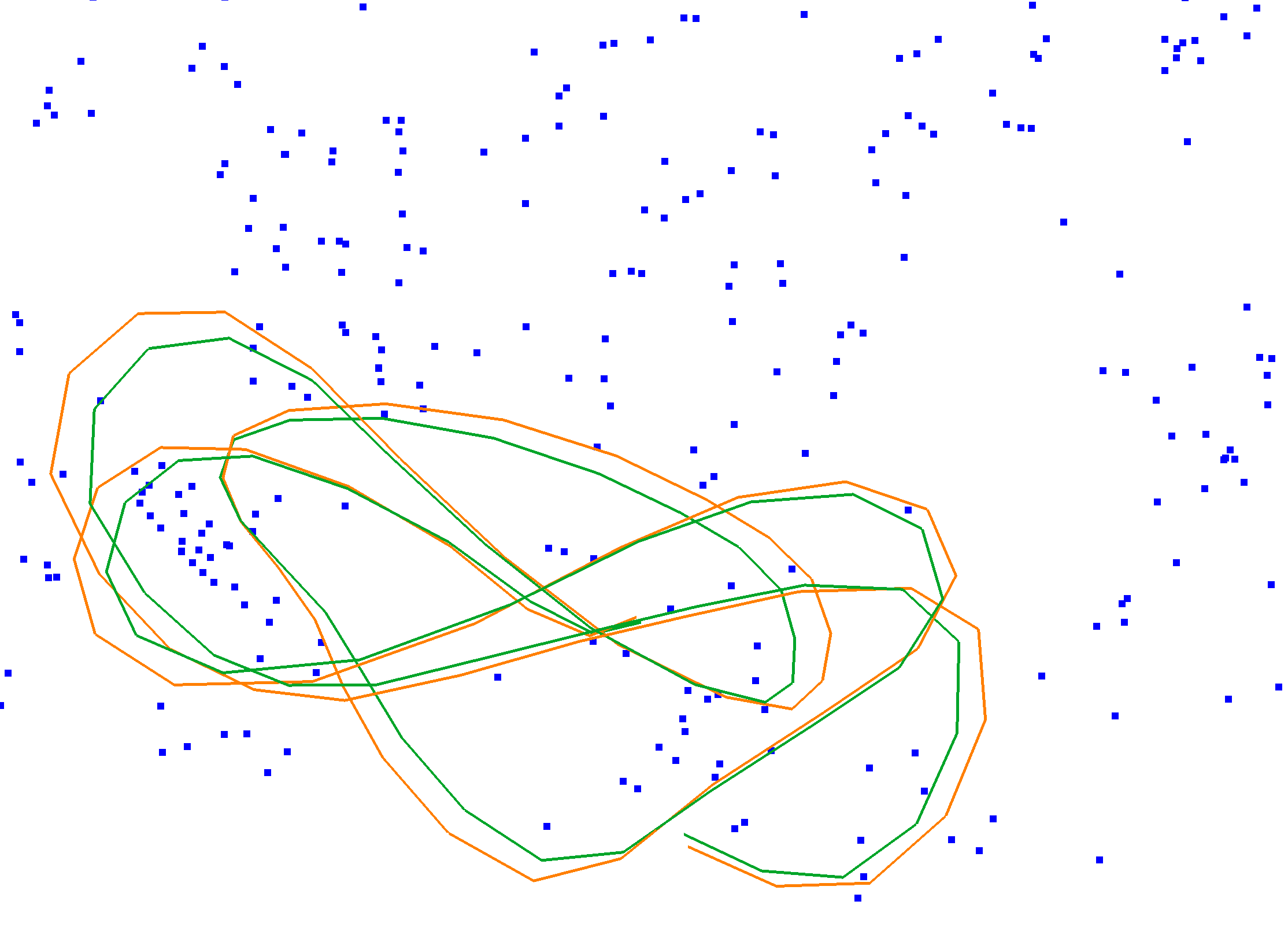}
		\captionsetup{justification=centering}
		\caption{Office sequence\\$\text{\ac{rmse}}=6.599~\text{mm}$}
		\label{fig:trajectories-c}
	\end{subfigure}
	\caption{Difference between the trajectories of the ground truth (\textit{in orange}) and the camera poses estimated during calibration (\textit{in green}) for the three sequences.}
	\label{fig:trajectories}
\end{figure}

In all previous experiments, the camera model provided by LiFCal is compared to the reference calibration.
While the estimated parameters demonstrate high reproducibility, the experiments would not be able to disclose potential biases caused by the camera model.
Therefore, we compare the camera poses obtained during the plenoptic bundle adjustment of the LiFCal calibration to the ground truth poses of~\cite{zeller2018dataset} obtained with a synchronized stereo camera.
Since the compared poses are obtained by two different sensors, potential model biases can be revealed.
\cref{fig:trajectories} shows the camera trajectory of the stereo reference from~\cite{zeller2018dataset} and the one obtained by LiFCal, with the \ac{rmse} for each sequence.
Positions are determined with centimeter-level precision for movements around 1~m in magnitude.
The slightly higher error in the Hallway sequence can be attributed to the larger scale of the scene and the increased distance between objects and the camera.

\subsubsection{Generalization to a Different Camera.}

Previous experiments were all performed with the same camera.
To demonstrate the generalization of our pipeline to a different sensor, we also used a Raytrix R25 camera (model: R25-C-D-10G-A018-A), with a resolution of 5320~pixels~$\times$~4600~pixels, a nominal focal length of $f_L = 12$~mm and a pixel size of 2.74~µm.
Additional camera acquisitions introduce Aruco markers~\cite{garrido2014automatic} printed on a sheet with known identifications and spacing. The markers are only used to define the scale of the scene directly.
The acquired sequences are: Table (42 images), Phone (51 images), Keyboard (50 images). See supplementary material for reference.

\setlength{\tabcolsep}{9pt}
\begin{table}[tb]
	\centering
	\caption{Estimated R25 camera parameters on three scenes scaled with Aruco markers and their \ac{sd}, to assess calibration repeatability across sequences. \ac{sd} values are expressed as percentage error relative to the estimated mean parameter value for consistent analysis regardless of parameter magnitude.}
	\label{tab:calibrationMarkers}
	\begin{tabular}{c | c c c c c}
		\toprule
		Scene & $f_L$ [mm] & $b_{L0}$ [mm] & $B$ [mm] & $c_x$ [pixel] & $c_y$ [pixel] \\
		\midrule
		Table & 13.163 & 11.893 & 0.386 & 2681.9 & 2272.0\\
		Phone & 13.062 & 11.927 & 0.330 & 2683.5 & 2297.5\\
		Keyboard & 12.977 & 11.929 & 0.310 & 2688.5 & 2295.4 \\
		\bottomrule
		SD [\%] & 0.713 & 0.170 & 11.519 & 0.128 & 0.619 \\
		\bottomrule
	\end{tabular}
\end{table}

\cref{tab:calibrationMarkers} shows results from sequences with the same camera configuration. Estimated parameters are consistent across experiments, with a relative \ac{sd} below 0.713\% except for parameter $B$ (\ac{sd} of 11.519\%) for the same reason as previously noted. The absence of ground truth prevents defining the \ac{rmse} for the R25 camera.

\subsection{Depth Estimation and SLAM}
\label{sec:downstreamTasks}

To showcase the usability and accuracy of the camera model obtained by LiFCal, we assess its effectiveness through various downstream tasks.

\subsubsection{Metric Depth Estimation.}
\begin{figure}[t]
	\centering
		\begin{subfigure}{0.24\linewidth}
			\includegraphics[width=1.0\linewidth]{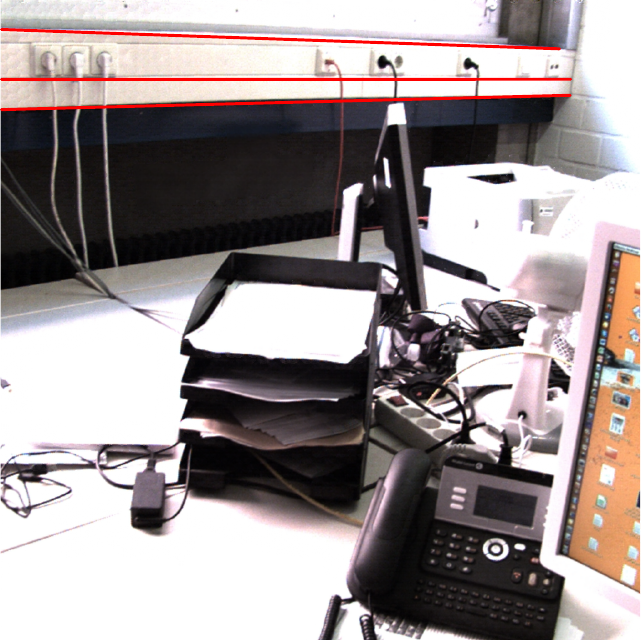}
			\caption{Image uncalibrated}
			\label{fig:distortionCorrection-a}
		\end{subfigure}
		\hfill
		\begin{subfigure}{0.24\linewidth}
			\includegraphics[width=1.0\linewidth]{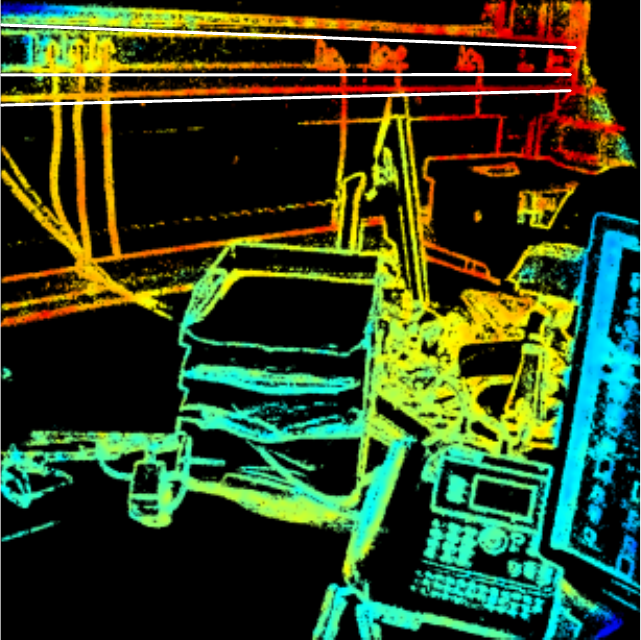}
			\caption{Depth uncalibrated}
			\label{fig:distortionCorrection-b}
		\end{subfigure}
		\begin{subfigure}{0.24\linewidth}
			\includegraphics[width=1.0\linewidth]{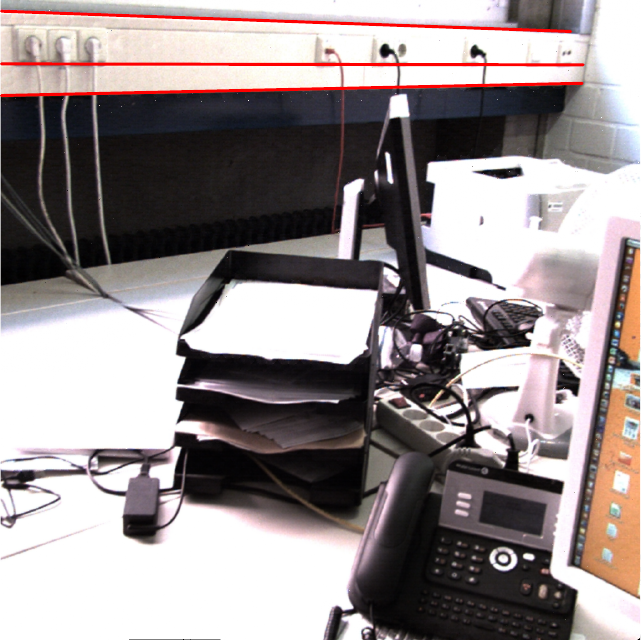}
			\caption{Image calibrated}
			\label{fig:distortionCorrection-c}
		\end{subfigure}
		\hfill
		\begin{subfigure}{0.24\linewidth}
			\includegraphics[width=1.0\linewidth]{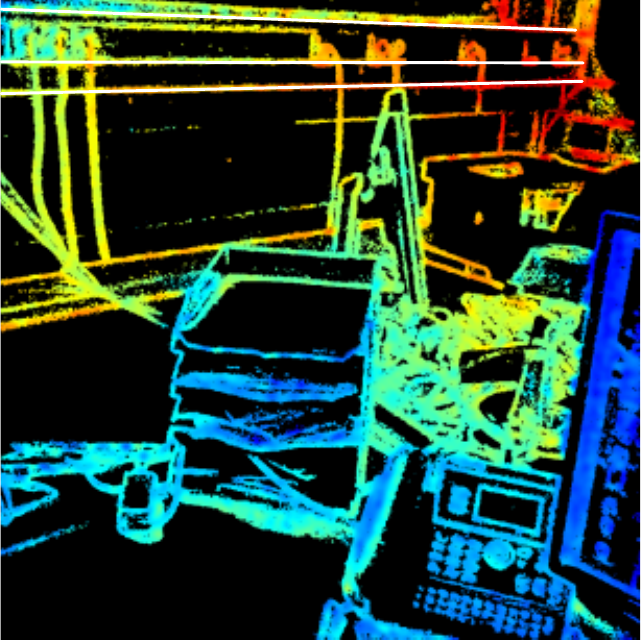}
			\caption{Depth calibrated}
			\label{fig:distortionCorrection-d}
		\end{subfigure}
	\caption{Distortion correction in a totally focused image and a depth map: (\protect\subref{fig:distortionCorrection-a}) and (\protect\subref{fig:distortionCorrection-b})  show an uncalibrated image and its depth map with non-straight lines (respectively \textit{red} and \textit{white}); (\protect\subref{fig:distortionCorrection-c}) and (\protect\subref{fig:distortionCorrection-d}) show a calibrated image with its depth map of the same scene with straight lines.}
	\label{fig:distortionCorrection}

\centering
\begin{minipage}[b]{.55\textwidth}
\setlength{\tabcolsep}{2pt}
    \begin{table}[H]
      \centering
      \caption{\ac{rmse} and scene scale calculated with ORB-SLAM3 and the calibrated plenoptic camera data against ground truth supplied by~\cite{zeller2018dataset}.}
      \label{tab:slamResults}
      \begin{tabular}{c | c c c c c}
        \toprule
        Metric & Lab & Hallway & Office & Parking \\
        \midrule
        \ac{rmse} [mm] & 9.775 & 78.993 & 6.060 & 46.636 \\
        Scale & 0.980 & 0.847 & 0.999 & 1.057 \\
        \bottomrule
      \end{tabular}
    \end{table}
\end{minipage}\hfill
\begin{minipage}[b]{.44\textwidth}
    \begin{figure}[H]
        \centering
        \includegraphics[width=0.9\linewidth]{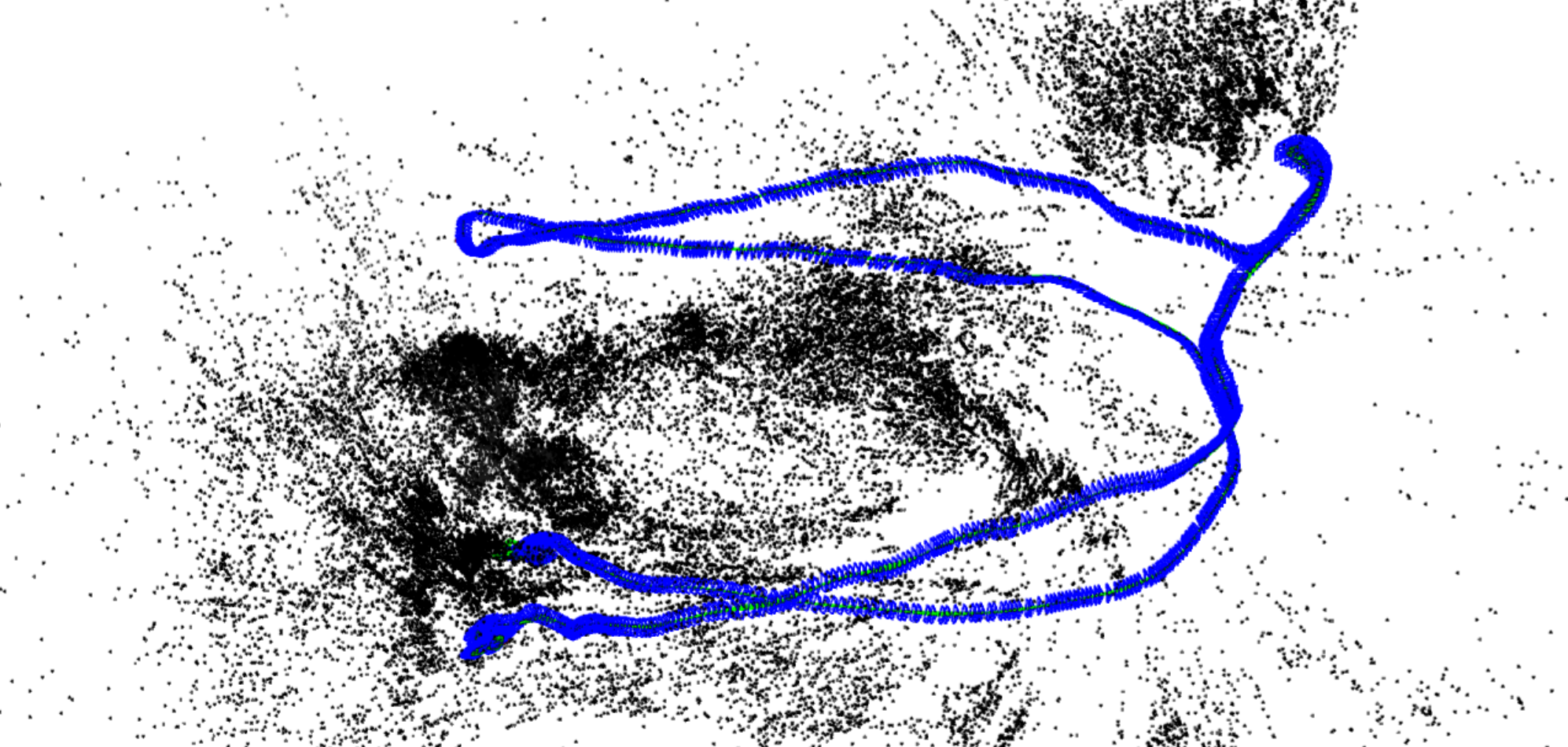}
        
        \caption{SLAM with calibrated plenoptic data on the Office sequence of~\cite{zeller2018dataset}.}
        \label{fig:SLAMOffice}
    \end{figure}
\end{minipage}
\end{figure}

Unlike other plenoptic camera models~\cite{heinze2016automated,zeller2014plencal}, LiFCal incorporates lens distortion directly into the raw image.
This allows distortion to be corrected during the depth estimation and totally focused image synthesis.
We extend an existing depth estimation pipeline~\cite{zeller2015plenopticdepth,zeller2015filtering} to directly generate undistorted metric depth maps and totally focused images (see \cref{fig:processOverview}) by using the camera model obtained by LiFCal.
As shown in \cref{fig:distortionCorrection}, the plenoptic camera model applied to the raw images prevents lens distortion from both the totally focused image (compare \ref{fig:distortionCorrection-a} and \ref{fig:distortionCorrection-c}) and the estimated depth (compare \ref{fig:distortionCorrection-b} and \ref{fig:distortionCorrection-d}).
Also, as shown in \cref{fig:processOverview}, an estimated depth map can be projected to a metric 3D point cloud (more examples in the supplementary material).

\subsubsection{Plenoptic-Camera-based SLAM.}
We calculate undistorted totally focused images and metric depth maps for sequences of~\cite{zeller2018dataset}.
Here, we project the virtual image space to a common image plane (using a central perspective projection explained in the supplementary material) to mimic an RGB-D sensor based on a pinhole camera model.
Using this data, we run ORB-SLAM3~\cite{campos2021orb} in RGB-D mode without loop closure.
\cref{tab:slamResults} reports the trajectory error compared to reference poses and scale for the winding part of both indoors and outdoors scenes of~\cite{zeller2018dataset}. The \ac{rmse} is below one centimeter for the Lab and Office scenes and slightly higher for Hallway and Parking due to larger displacements and distances.
Notably, plenoptic camera depth data achieves accurate scene scaling, comparable to a dedicated plenoptic \ac{slam} approach~\cite{zeller2018spo}.
For example, \cref{fig:SLAMOffice} illustrates estimated trajectory for the Office sequence (seq. 9 of~\cite{zeller2018dataset}), achieving an absolute scale error of 1.1\%, a scale drift of 2.3\%, a translational alignment error of 0.9\% and a rotational error of 18.4$^\circ$ accumulated over the entire trajectory. Therefore, except for the rotational error, results for this sequence are similar to the ones of~\cite{zeller2018spo}, despite ORB-SLAM3 not being optimized for plenoptic camera data.

\section{Conclusion}
\label{sec:conclusion}

We introduce LiFCal, a new online calibration pipeline for \ac{mla}-based light field cameras.
We demonstrate that LiFCal can determine accurate and repeatable calibration results on target-free scenes with sufficient camera motion and features.
By fixing camera specific pre-calibrated parameters, LiFCal can obtain accurate metric parameters even without any prior knowledge about the scene scale.
We integrated the plenoptic camera model into a depth estimation pipeline to obtain metric point clouds of the scene.
Besides metric depth estimation, we demonstrate the usability of the obtained camera model  by integrating the calculated depth maps and totally focused images into a state-of-the-art RGB-D \ac{slam} system.

{\small \subsubsection{Acknowledgments.} We would like to thank the company Raytrix for providing a light field camera, which enabled us to conduct the experiments with data obtained by ourselves.}

%
%
%
%
\bibliographystyle{splncs04}
\bibliography{egbib}

\begin{thebibliography}{10}
\providecommand{\url}[1]{\texttt{#1}}
\providecommand{\urlprefix}{URL }
\providecommand{\doi}[1]{https://doi.org/#1}

\bibitem{adelson1992single}
Adelson, E.H., Wang, J.Y.: Single lens stereo with a plenoptic camera. Transactions on Pattern Analysis and Machine Intelligence (TPAMI)  \textbf{14}(2),  99--106 (1992). \doi{10.1109/34.121783}

\bibitem{Agarwal_Ceres_Solver_2022}
Agarwal, S., Mierle, K., Team, T.C.S.: Ceres solver (Oct 2023)

\bibitem{bok2017geometric}
Bok, Y., Jeon, H.G., Kweon, I.S.: Geometric calibration of micro-lens-based light field cameras using line features. Transactions on Pattern Analysis and Machine Intelligence (TPAMI)  \textbf{39}(2),  287--300 (2017). \doi{10.1109/tpami.2016.2541145}

\bibitem{Brown1966DecenteringDO}
Brown, D.: Decentering distortion of lenses. Photogrammetric Engineering  \textbf{32}(3),  444--462 (1966)

\bibitem{campos2021orb}
Campos, C., Elvira, R., Rodr{\'\i}guez, J.J.G., Montiel, J.M., Tard{\'o}s, J.D.: Orb-slam3: An accurate open-source library for visual, visual–inertial, and multimap slam. Transactions on Robotics (T-RO)  \textbf{37}(6),  1874--1890 (2021). \doi{10.1109/tro.2021.3075644}

\bibitem{caprile1990using}
Caprile, B., Torre, V.: Using vanishing points for camera calibration. International journal of computer vision (IJCV)  \textbf{4}(2),  127--139 (1990). \doi{10.1007/BF00127813}

\bibitem{chang2024target}
Chang, D., Huang, S., Zhou, Y., Qin, X., Ding, R., Hu, M.: Target-free stereo camera-gnss/imu self-calibration based on iterative refinement. Sensors Journal  \textbf{24}(3),  3722–3730 (2024). \doi{10.1109/jsen.2023.3343371}

\bibitem{cho2013modeling}
Cho, D., Lee, M., Kim, S., Tai, Y.W.: Modeling the calibration pipeline of the lytro camera for high quality light-field image reconstruction. In: International Conference on Computer Vision (ICCV). pp. 3280--3287. IEEE (2013). \doi{10.1109/iccv.2013.407}

\bibitem{dansereau2013decoding}
Dansereau, D.G., Pizarro, O., Williams, S.B.: Decoding, calibration and rectification for lenselet-based plenoptic cameras. In: Conference on Computer Vision and Pattern Recognition (CVPR). pp. 1027--1034. IEEE (2013). \doi{10.1109/cvpr.2013.137}

\bibitem{darwish2019plenoptic}
Darwish, W., Bolsee, Q., Munteanu, A.: Plenoptic camera calibration based on sub-aperture images. In: International Conference on Image Processing (ICIP). pp. 3527--3531. IEEE (2019). \doi{10.1109/icip.2019.8803473}

\bibitem{engel2018direct}
Engel, J., Koltun, V., Cremers, D.: Direct sparse odometry. Transactions on Pattern Analysis and Machine Intelligence (TPAMI)  \textbf{40}(3),  611--625 (2018). \doi{10.1109/TPAMI.2017.2658577}

\bibitem{engel2014lsd}
Engel, J., Sch{\"o}ps, T., Cremers, D.: Lsd-slam: Large-scale direct monocular slam. In: European Conference on Computer Vision (ECCV). pp. 834--849. Springer (2014). \doi{10.1007/978-3-319-10605-2\_54}

\bibitem{fachada2022pattern}
Fachada, S., Bonatto, D., Losfeld, A., Lafruit, G., Teratani, M.: Pattern-free plenoptic 2.0 camera calibration. In: International Workshop on Multimedia Signal Processing (MMSP). pp.~1--6. IEEE (2022). \doi{10.1109/mmsp55362.2022.9949312}

\bibitem{fachada2021calibration}
Fachada, S., Losfeld, A., Senoh, T., Lafruit, G., Teratani, M.: A calibration method for subaperture views of plenoptic 2.0 camera arrays. In: International Workshop on Multimedia Signal Processing (MMSP). pp.~1--6. IEEE (2021). \doi{10.1109/mmsp53017.2021.9733556}

\bibitem{fang2022self}
Fang, J., Vasiljevic, I., Guizilini, V., Ambrus, R., Shakhnarovich, G., Gaidon, A., Walter, M.R.: Self-supervised camera self-calibration from video. In: International Conference on Robotics and Automation (ICRA). pp. 8468--8475. IEEE (2022). \doi{10.1109/icra46639.2022.9811784}

\bibitem{fischler1981random}
Fischler, M.A., Bolles, R.C.: Random sample consensus: A paradigm for model fitting with applications to image analysis and automated cartography. Communications of the ACM  \textbf{24}(6),  381--395 (Jun 1981). \doi{10.1145/358669.358692}

\bibitem{garrido2014automatic}
Garrido-Jurado, S., Mu{\~n}oz-Salinas, R., Madrid-Cuevas, F.J., Mar{\'\i}n-Jim{\'e}nez, M.J.: Automatic generation and detection of highly reliable fiducial markers under occlusion. Pattern Recognition  \textbf{47}(6),  2280--2292 (2014). \doi{10.1016/j.patcog.2014.01.005}

\bibitem{heinze2016automated}
Heinze, C., Spyropoulos, S., Hussmann, S., Perwa{\ss}, C.: Automated robust metric calibration algorithm for multifocus plenoptic cameras. Transactions on Instrumentation and Measurement (TIM)  \textbf{65}(5),  1197--1205 (2016). \doi{10.1109/tim.2015.2507412}

\bibitem{ives1903parallax}
Ives, F.E.: Parallax stereogram and process of making same (1903), uS Patent 725,567

\bibitem{ji2016light}
Ji, Z., Zhang, C., Wang, Q.: Light field camera self-calibration and registration. In: Optoelectronic Imaging and Multimedia Technology. vol. 10020, pp. 56--65. SPIE (2016). \doi{10.1117/12.2246339}

\bibitem{johannsen2013calibration}
Johannsen, O., Heinze, C., Goldluecke, B., Perwa{\ss}, C.: On the calibration of focused plenoptic cameras. In: Time-of-Flight and Depth Imaging. Sensors, Algorithms, and Applications: Dagstuhl 2012 Seminar on Time-of-Flight Imaging and GCPR 2013 Workshop on Imaging New Modalities. pp. 302--317. Springer (2013). \doi{10.1007/978-3-642-44964-2\_15}

\bibitem{labussiere2020blur}
Labussi{\`e}re, M., Teuli{\`e}re, C., Bernardin, F., Ait-Aider, O.: Blur aware calibration of multi-focus plenoptic camera. In: Conference on Computer Vision and Pattern Recognition (CVPR). pp. 2542--2551. IEEE (2020). \doi{10.1109/cvpr42600.2020.00262}

\bibitem{labussiere2022leveraging}
Labussi{\`e}re, M., Teuli{\`e}re, C., Bernardin, F., Ait-Aider, O.: Leveraging blur information for plenoptic camera calibration. International journal of computer vision (IJCV)  \textbf{130}(7),  1655--1677 (2022). \doi{10.1007/s11263-022-01582-z}

\bibitem{lippmann1908epreuves}
Lippmann, G.: Epreuves reversibles donnant la sensation du relief. Journal de Physique Théorique et Appliquée  \textbf{7}(1),  821--825 (1908). \doi{10.1051/jphystap:019080070082100}

\bibitem{lopez2019deep}
Lopez, M., Mari, R., Gargallo, P., Kuang, Y., Gonzalez-Jimenez, J., Haro, G.: Deep single image camera calibration with radial distortion. In: Conference on Computer Vision and Pattern Recognition (CVPR). pp. 11809--11817. IEEE (2019). \doi{10.1109/cvpr.2019.01209}

\bibitem{lumsdaine2009focused}
Lumsdaine, A., Georgiev, T.: The focused plenoptic camera. In: International Conference on Computational Photography (ICCP). pp.~1--8. IEEE (2009). \doi{10.1109/iccphot.2009.5559008}

\bibitem{lumsdaine2008full}
Lumsdaine, A., Georgiev, T., et~al.: Full resolution lightfield rendering. Indiana University and Adobe Systems, Tech. Rep  \textbf{91}, ~92 (2008)

\bibitem{mur2017orb}
Mur-Artal, R., Tard{\'o}s, J.D.: Orb-slam2: An open-source slam system for monocular, stereo, and rgb-d cameras. Transactions on Robotics (T-RO)  \textbf{33}(5),  1255--1262 (2017). \doi{10.1109/tro.2017.2705103}

\bibitem{ng2006digital}
Ng, R.: Digital Light Field Photography. Stanford University (2006)

\bibitem{ng2005light}
Ng, R., Levoy, M., Br{\'e}dif, M., Duval, G., Horowitz, M., Hanrahan, P.: Light Field Photography with a Hand-held Plenoptic Camera. Ph.D. thesis, Stanford University (2005)

\bibitem{o2018calibrating}
O'brien, S., Trumpf, J., Ila, V., Mahony, R.: Calibrating light-field cameras using plenoptic disc features. In: International conference on 3D vision (3DV). pp. 286--294. IEEE (2018). \doi{10.1109/3dv.2018.00041}

\bibitem{perwass2012single}
Perwass, C., Wietzke, L.: Single lens 3d-camera with extended depth-of-field. In: Human Vision and Electronic Imaging (HVEI). vol.~8291, pp. 45--59. SPIE (2012). \doi{10.1117/12.909882}

\bibitem{pollefeys1997stratified}
Pollefeys, M., Van~Gool, L.: A stratified approach to metric self-calibration. In: Conference on Computer Vision and Pattern Recognition (CVPR). pp. 407--412. IEEE (1997). \doi{10.1109/cvpr.1997.609357}

\bibitem{quenzel2017online}
Quenzel, J., Rosu, R.A., Houben, S., Behnke, S.: Online depth calibration for rgb-d cameras using visual slam. In: International Conference on Intelligent Robots and Systems (IROS). pp. 2227--2234. IEEE (2017). \doi{10.1109/iros.2017.8206043}

\bibitem{rehder2017online}
Rehder, E., Kinzig, C., Bender, P., Lauer, M.: Online stereo camera calibration from scratch. In: Intelligent Vehicles Symposium (IV). pp. 1694--1699. IEEE (2017). \doi{10.1109/ivs.2017.7995952}

\bibitem{schoenberger2016sfm}
Sch\"{o}nberger, J.L., Frahm, J.M.: Structure-from-motion revisited. In: Conference on Computer Vision and Pattern Recognition (CVPR). pp. 4104--4113. IEEE (2016). \doi{10.1109/CVPR.2016.445}

\bibitem{wang2017stereo}
Wang, R., Schworer, M., Cremers, D.: Stereo dso: Large-scale direct sparse visual odometry with stereo cameras. In: International Conference on Computer Vision (ICCV). pp. 3903--3911. IEEE (2017). \doi{10.1109/iccv.2017.421}

\bibitem{wang2022occlusion}
Wang, Y., Wang, L., Liang, Z., Yang, J., An, W., Guo, Y.: Occlusion-aware cost constructor for light field depth estimation. In: Conference on Computer Vision and Pattern Recognition (CVPR). pp. 19777--19786. IEEE (2022). \doi{10.1109/cvpr52688.2022.01919}

\bibitem{xiao2023cutmib}
Xiao, Z., Liu, Y., Gao, R., Xiong, Z.: Cutmib: Boosting light field super-resolution via multi-view image blending. In: Conference on Computer Vision and Pattern Recognition (CVPR). pp. 1672--1682. IEEE (2023). \doi{10.1109/cvpr52729.2023.00167}

\bibitem{zeisl2016structure}
Zeisl, B., Pollefeys, M.: Structure-based auto-calibration of rgb-d sensors. In: International Conference on Robotics and Automation (ICRA). pp. 5076--5083. IEEE (2016). \doi{10.1109/icra.2016.7487713}

\bibitem{zeller2017plencalib}
Zeller, N., Noury, C.A., Quint, F., Teulière, C., Stilla, U., Dhome, M.: Metric calibration of a focused plenoptic camera based on a 3d calibration target. ISPRS Annals of the Photogrammetry, Remote Sensing and Spatial Information Sciences (ISPRS Annals)  \textbf{III–3},  449–456 (2016). \doi{10.5194/isprsannals-iii-3-449-2016}

\bibitem{zeller2018dissertation}
Zeller, N.: Direct Plenoptic Odometry -- Robust Tracking and Mapping with a Light Field Camera. Ph.D. thesis, Technische Universit{\"a}t M{\"u}nchen (2018)

\bibitem{zeller2014plencal}
Zeller, N., Quint, F., Stilla, U.: Calibration and accuracy analysis of a focused plenoptic camera. ISPRS Annals of the Photogrammetry, Remote Sensing and Spatial Information Sciences (ISPRS Annals)  \textbf{II–3},  205–212 (2014). \doi{10.5194/isprsannals-ii-3-205-2014}

\bibitem{zeller2015plenopticdepth}
Zeller, N., Quint, F., Stilla, U.: Establishing a probabilistic depth map from focused plenoptic cameras. In: International Conference on 3D Vision (3DV). pp. 91--99. IEEE (2015). \doi{10.1109/3dv.2015.18}

\bibitem{zeller2015filtering}
Zeller, N., Quint, F., Stilla, U.: Filtering probabilistic depth maps received from a focused plenoptic camera. BW-CAR Symposium on Information and Communication Systems (SInCom)  \textbf{2},  7--12 (2015)

\bibitem{zeller2016depthcalib}
Zeller, N., Quint, F., Stilla, U.: Depth estimation and camera calibration of a focused plenoptic camera for visual odometry. ISPRS Journal of Photogrammetry and Remote Sensing (P\&RS)  \textbf{118},  83–100 (2016). \doi{10.1016/j.isprsjprs.2016.04.010}

\bibitem{zeller2017dpo}
Zeller, N., Quint, F., Stilla, U.: From the calibration of a light-field camera to direct plenoptic odometry. Journal of Selected Topics in Signal Processing  \textbf{11}(7),  1004–1019 (2017). \doi{10.1109/jstsp.2017.2737965}

\bibitem{zeller2018spo}
Zeller, N., Quint, F., Stilla, U.: Scale-awareness of light field camera based visual odometry. In: European Conference on Computer Vision (ECCV). p. 732–747. Springer (2018). \doi{10.1007/978-3-030-01237-3\_44}

\bibitem{zeller2018dataset}
Zeller, N., Quint, F., Stilla, U.: A synchronized stereo and plenoptic visual odometry dataset. arXiv preprint  (2018). \doi{10.48550/arXiv.1807.09372}

\bibitem{zhao2020metric}
Zhao, Y., Li, H., Mei, D., Shi, S.: Metric calibration of unfocused plenoptic cameras for three-dimensional shape measurement. Optical Engineering  \textbf{59}(7),  073104--073104 (2020). \doi{10.1117/1.oe.59.7.073104}

\bibitem{zhou2019two}
Zhou, P., Cai, W., Yu, Y., Zhang, Y., Zhou, G.: A two-step calibration method of lenslet-based light field cameras. Optics and Lasers in Engineering  \textbf{115},  190--196 (2019). \doi{10.1016/j.optlaseng.2018.11.024}

\end{thebibliography}


\clearpage

\title{LiFCal: Online Light Field Camera Calibration via Bundle Adjustment\\Supplementary Material}

\ifreview
\else
	\titlerunning{LiFCal -- Supplementary Material}

	\author{Aymeric Fleith\inst{1,2} \and
	Doaa Ahmed\inst{2} \and
	Daniel Cremers\inst{1} \and
	Niclas Zeller\inst{2}}
	
	\authorrunning{A. Fleith et al.}
	
	\institute{Technical University of Munich, Munich, Germany\\
    \email{\{aymeric.fleith, cremers\}@tum.de}\\ 
    \and Karlsruhe University of Applied Sciences, Karlsruhe, Germany\\
	\email{\{doaa.ahmed, niclas.zeller\}@h-ka.de}}
\fi

\maketitle              

\renewcommand*{\thesection}{\Alph{section}}
\section{Introduction}
\label{sec:introduction}

This supplementary material provides additional details and results beyond those in the main paper.
Specifically, this includes the results of metric point clouds obtained using LiFCal calibration (\cref{sec:metricDepthMapResults}), the calculation of RGB-D data used for a \ac{slam} task to mimic the data of a pinhole model (\cref{sec:RGBDDataORBSLAM}), a detailed definition of the implemented distortion model (\cref{sec:distorsionModel}), additional details regarding the initialization of the plenoptic camera model to perform bundle adjustment (\cref{sec:parameterInitialization}), sample images of the sequences used for the calibration experiments (\cref{sec:SsequencesExperiments}).

\section{Metric depth map results}
\label{sec:metricDepthMapResults}

To demonstrate the usability and accuracy of the LiFCal calibration method, we present results on several downstream tasks (see Sec. 4.2 in the main paper). This section presents examples of metric point clouds obtained by applying LiFCal calibration to raw data from several scenes.

\cref{fig:metricDepthMapResults} shows the resulting metric point clouds. For each scene, the following images are provided: raw image from the plenoptic camera, totally focused image corrected by the camera model obtained by LiFCal, depth map corrected by the camera model obtained by LiFCal, metric point cloud. Two calibration processes evaluated in the experiments were used in order to demonstrate performance of both. For the scenes in \cref{fig:metricDepthMapResults-a}, \cref{fig:metricDepthMapResults-b}, \cref{fig:metricDepthMapResults-c}, calibration is performed using the scene from \cref{fig:metricDepthMapResults-a}. The calculated model is retained for the sequences in \cref{fig:metricDepthMapResults-b} and \cref{fig:metricDepthMapResults-c}. For the other sequences in \cref{fig:metricDepthMapResults}, camera calibration is performed with markers and kept fixed for sequences in \cref{fig:metricDepthMapResults-d}, \cref{fig:metricDepthMapResults-e}, \cref{fig:metricDepthMapResults-f}, \cref{fig:metricDepthMapResults-g}, \cref{fig:metricDepthMapResults-h}, \cref{fig:metricDepthMapResults-i}. In the right-hand column, the point cloud is completed with a distance scale and a depth scale, both expressed in millimeters.

\begin{figure}[H]
  \centering
  \adjustbox{minipage=1.3em,valign=t}{}
  \begin{subfigure}[t,valign=t]{\dimexpr1.0\linewidth-1.3em\relax}
  \hspace{13.2mm}
  \begin{tikzpicture}
\matrix [matrix of nodes,column sep=13.1mm,row sep=1mm,inner sep=0mm]
{%
\node(i00){\makecell{Raw \\ image}}; &
\node(i02){\makecell{Totally focused \\ image}};  &
\node(i06){\makecell{Depth \\ map}};  &
\node(i08){\makecell{Metric \\ point cloud}};\\
};
\end{tikzpicture}
  \end{subfigure}
  \vfill \vspace{2mm}
  \adjustbox{minipage=1.3em,valign=t}{\subcaption{}\label{fig:metricDepthMapResults-a}}%
  \begin{subfigure}[t,valign=t]{\dimexpr1.0\linewidth-1.3em\relax}
  \centering
  \input{images/Metric_depth_map/desktop/desktop}
  \end{subfigure}
  \vfill \vspace{2mm}
  \adjustbox{minipage=1.3em,valign=t}{\subcaption{}\label{fig:metricDepthMapResults-b}}%
  \begin{subfigure}[t,valign=t]{\dimexpr1.0\linewidth-1.3em\relax}
  \centering
  \input{images/Metric_depth_map/tennis_balls/tennis_balls}
  \end{subfigure}
  \vfill \vspace{2mm}
  \adjustbox{minipage=1.3em,valign=t}{\subcaption{}\label{fig:metricDepthMapResults-c}}%
  \begin{subfigure}[t,valign=t]{\dimexpr1.0\linewidth-1.3em\relax}
  \centering
  \input{images/Metric_depth_map/helmet/helmet}
  \end{subfigure}
  \vfill \vspace{2mm}
  \adjustbox{minipage=1.3em,valign=t}{\subcaption{}\label{fig:metricDepthMapResults-d}}%
  \begin{subfigure}[t,valign=t]{\dimexpr1.0\linewidth-1.3em\relax}
  \centering
  \input{images/Metric_depth_map/car_bikes/car_bikes}
  \end{subfigure}
  \vfill \vspace{2mm}
  \adjustbox{minipage=1.3em,valign=t}{\subcaption{}\label{fig:metricDepthMapResults-e}}%
  \begin{subfigure}[t,valign=t]{\dimexpr1.0\linewidth-1.3em\relax}
  \centering
  \input{images/Metric_depth_map/knights/knights}
  \end{subfigure}
  \vfill \vspace{2mm}
  \adjustbox{minipage=1.3em,valign=t}{\subcaption{}\label{fig:metricDepthMapResults-f}}%
  \begin{subfigure}[t,valign=t]{\dimexpr1.0\linewidth-1.3em\relax}
  \centering
  \input{images/Metric_depth_map/squirrel/squirrel}
  \end{subfigure}
  \vfill \vspace{2mm}
\end{figure}
\begin{figure}[H]\ContinuedFloat
    \centering
  \adjustbox{minipage=1.3em,valign=t}{\subcaption{}\label{fig:metricDepthMapResults-g}}%
  \begin{subfigure}[t,valign=t]{\dimexpr1.0\linewidth-1.3em\relax}
  \centering
  \input{images/Metric_depth_map/gameboy/gameboy}
  \end{subfigure}
  \vfill \vspace{2mm}
  \adjustbox{minipage=1.3em,valign=t}{\subcaption{}\label{fig:metricDepthMapResults-h}}%
  \begin{subfigure}[t,valign=t]{\dimexpr1.0\linewidth-1.3em\relax}
  \centering
  \input{images/Metric_depth_map/workstation/workstation}
  \end{subfigure}
  \vfill \vspace{2mm}
  \adjustbox{minipage=1.3em,valign=t}{\subcaption{}\label{fig:metricDepthMapResults-i}}%
  \begin{subfigure}[t,valign=t]{\dimexpr1.0\linewidth-1.3em\relax}
  \centering
  \input{images/Metric_depth_map/office/office}
  \end{subfigure}%
   \caption{Depth map results for several scenes. The point clouds in (\protect\subref{fig:metricDepthMapResults-a}), (\protect\subref{fig:metricDepthMapResults-b}), (\protect\subref{fig:metricDepthMapResults-c}) are generated with a calibration on the scene (\protect\subref{fig:metricDepthMapResults-a}). The point clouds in (\protect\subref{fig:metricDepthMapResults-d}), (\protect\subref{fig:metricDepthMapResults-e}), (\protect\subref{fig:metricDepthMapResults-f}), (\protect\subref{fig:metricDepthMapResults-g}), (\protect\subref{fig:metricDepthMapResults-h}), (\protect\subref{fig:metricDepthMapResults-i}) are generated with a calibration based on markers. In order, the different columns show: the raw image from the plenoptic camera, the totally focused image corrected by the LiFCal calibration, the depth map corrected by the LiFCal calibration and the metric point cloud of the scene. The point clouds are accompanied by a distance scale (\textit{bottom right of each point cloud}) and a depth scale (\textit{right of each point cloud}). Distance and depth data in the point clouds are in millimeters.}
   \label{fig:metricDepthMapResults}
\end{figure}

\section{RGB-D data calculation to run ORB-SLAM}
\label{sec:RGBDDataORBSLAM}

We demonstrate the effectiveness of LiFCal calibration by integrating it into a \ac{slam} task using the plenoptic camera (see Sec. 4.2 in the main paper). For this, we use ORB-SLAM3~\cite{campos2021orb} in its version with RGB-D data, disabling loop closure.

\begin{figure}[t]
  \centering
   \def\svgwidth{1.0\linewidth}
   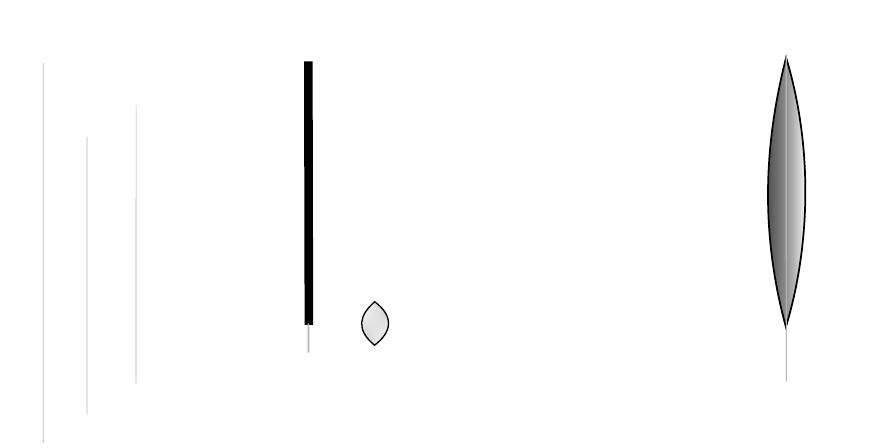

   \caption{Central perspective projection of the virtual image to a common image plane to mimic a pinhole camera model to use the data with ORB-SLAM. The points of the virtual image are projected along a straight line passing through the center of the main lens (\textit{blue}) instead of horizontally (\textit{orange}). The distance between the \ac{mla} and the projection plane is chosen at $2B$.}
   \label{fig:centralPerspectiveProjection}
\end{figure}

ORB-SLAM3 uses a pinhole camera model. To mimic this model with the plenoptic camera, we perform a central perspective projection as shown in \cref{fig:centralPerspectiveProjection}. Each point $X_V'$ in virtual space is formed at a different distance from the \ac{mla}, depending on its distance from the camera. Instead of projecting the points horizontally (as shown in orange), they are projected onto the projection plane along a straight line passing through the center of the main lens (illustrated projection in blue). In principle, the projection plane can be set at an arbitrary distance from the \ac{mla}. We choose a distance between the \ac{mla} and the projection plane of $2B$ corresponding to the total covering plane defined in~\cite{perwass2012single}. It corresponds to a virtual depth of $v=2$, which is the furthest distance from the camera for which a depth can be measured. This projection ensures that the edges of the image are not too large. Projection is performed using similar triangles. The coordinates of the new point $X_{proj}=[x_{proj}, y_{proj}]^T$ after projection of $X_V'=[x_V', y_V', z_V' = v]^T$ are calculated using \cref{eq:projX} and \cref{eq:projY}.
\begin{align}
  x_{proj} &= \frac{x_V' - c_x}{v \cdot B + b_{L0}} \cdot (2 \cdot B + b_{L0}) + c_x
  \label{eq:projX}\\
  y_{proj} &= \frac{y_V' - c_y}{v \cdot B + b_{L0}} \cdot (2 \cdot B + b_{L0}) + c_y
  \label{eq:projY}
\end{align}
In \cref{eq:projX} and \cref{eq:projY}, $v$ is the virtual depth of the point, $B$ is the distance between \ac{mla} and sensor, $b_{L0}$ is the distance between main lens and \ac{mla} and $C_L=[c_x,c_y]^T$ is the principal point of the main lens.

ORB-SLAM3 requires metric depth data. The metric depth is then determined by the thin lens equation using \cref{eq:depthPoint}.
The parameter $f_L$ is the focal length of the main lens, $b_L$ is the distance between the main lens and the virtual image and $z_C$ is the distance between the real object and the main lens.

\begin{align}
  z_C = \left(\frac{1}{f_L} - \frac{1}{b_L}\right)^{-1}  \quad \text{with}  \quad b_L = b + b_{L0}  \quad \text{and}  \quad b = v \cdot B
  \label{eq:depthPoint}
\end{align}

\section{Distortion Model}
\label{sec:distorsionModel}

As described in the main paper (see Sec. 3.1), lens distortion is defined directly on raw image coordinates $X_R=[x_R, y_R]^T$.
Using this model, we implicitly account for both main lens distortion, and sensor and \ac{mla} misalignment. The model uses radial symmetric distortion and tangential distortion according to the model presented in~\cite{Brown1966DecenteringDO}.
Nevertheless, it can be replaced by any other distortion model.
Distortion is applied both to the raw image points on the sensor and to the centers of the micro images or micro lenses respectively.
For the sake of notation, we define $x'_R$ and $y'_R$ as follows:
\begin{align}
   x'_R = x_R-c_x, \qquad y'_R = y_R-c_y.
  \label{eq:xyPrimeR}
\end{align}

The radial symmetric distortion is characterized by a polynomial with respect to the radius $r$ given in \cref{eq:DeltarPolynomial} where $k_n$ is the $(n+1)$-th coefficient. The radius $r$ is defined as the Euclidean distance between a point $X_R$ and the main lens principal point.
\begin{align}
  \Delta r_{rad} = \sum_{n=0}^{\infty} k_n r^{2n+3} \qquad \text{with} \qquad 
  r = \sqrt{{x'_R}^2 + {y'_R}^2}
  \label{eq:DeltarPolynomial}
\end{align}
Trigonometrically projecting \cref{eq:DeltarPolynomial} onto the two axes gives the $\Delta x_{rad}$ and $\Delta y_{rad}$ correction terms in the image's Cartesian coordinate system. Retaining only the first three coefficients results in \cref{eq:deltaXrad} and \cref{eq:deltaYrad} respectively.
\begin{align}
  \Delta x_{rad} &= x'_R \frac{\Delta r_{rad}}{r} = x'_R \cdot (k_0 r^2 + k_1 r^4 + k_2 r^6)
  \label{eq:deltaXrad}\\
  \Delta y_{rad} &= y'_R \frac{\Delta r_{rad}}{r} = y'_R \cdot (k_0 r^2 + k_1 r^4 + k_2 r^6)
  \label{eq:deltaYrad}
\end{align}

The tangential distortion is defined with the first two parameters $p_0$ and $p_1$. The expressions in both directions are given in \cref{eq:deltaXtan} and \cref{eq:deltaYtan} respectively.
\begin{align}
  \Delta x_{tan} &= p_0 \cdot \left(r^2 + 2 {x'_R}^2\right) + 2 p_1 x'_R y'_R
  \label{eq:deltaXtan}\\
  \Delta y_{tan} &= p_1 \cdot \left(r^2 + 2 {y'_R}^2\right) + 2 p_0 x'_R y'_R
  \label{eq:deltaYtan}
\end{align}
The radial and tangential corrections can be combined to obtain the coordinates of the distorted point $X_{Rd} = [x_{Rd}, y_{Rd}]^T$ from the base $x_R$ and $y_R$ coordinates:
\begin{align}
  x_{Rd} &= x_R + \Delta x_{rad} + \Delta x_{tan},
  \label{eq:xRD}\\
  y_{Rd} &= y_R + \Delta y_{rad} + \Delta y_{tan}.
  \label{eq:yRD}
\end{align}

Since the distortion model defined in \cref{eq:deltaXtan} and \cref{eq:deltaYtan} affects all raw image coordinates on the sensor, the same distortion model is implicitly applied to the micro image centers $C_I = [c_{Ix}, c_{Iy}]^T$ as well.

To project an object point onto the image sensor (\eg during plenoptic bundle adjustment), the distortion model can be applied in the forward direction as defined in \cref{eq:xRD} and \cref{eq:yRD}.
However, for downstream tasks like depth estimation, the distortion model needs to be inverted.
Because the defined polynomials are not directly invertible, this, in general, is done in an iterative manner.
Nevertheless, this undistortion process can be calculated beforehand and can be applied directly to the entire recorded raw image.
Defining the distortion on raw image coordinates has the advantage that after undistorting the raw image, tasks like depth estimation, image synthesis, etc. can be carried out without considering the distortion anymore.

\section{Initialization of plenoptic camera parameters}
\label{sec:parameterInitialization}

As described in the main paper (Sec. 3.2), after the initialization phase, the parameters of the plenoptic camera model need to be initialized.
The main lens focal length $f_L$ and the principal point $C_L$ can be set during the initialization using the pinhole model.
The additional plenoptic parameters $B$ and $b_{L0}$ are initialized by solving the linear \cref{eq:virtualDepth_supp} obtained from the plenoptic camera model.
This equation is set up for each feature point observed in a calibration image.
\begin{align}
	b_L = v\cdot B+b_{L0}
	\label{eq:virtualDepth_supp}
\end{align}
Here, $v$ is the virtual depth which can be estimated based on the recorded raw image \cite{zeller2015plenopticdepth} and $b_L$ is the corresponding main lens image distance which is obtained from the thin lens equation defined in \cref{eq:bL_supp}.
\begin{align}
	b_L=\left(\frac{1}{f_L}-\frac{1}{z_C}\right)^{-1}
	\label{eq:bL_supp}
\end{align}
In \cref{eq:bL_supp}, $z_C$ is the object distance \ie the third component of the camera coordinates $X_C = [x_C,y_C,z_C]^T$ of the corresponding object points.
For each point, $X_C$ is calculated using the estimation object point coordinates $X_W = [x_W, y_W, z_W]^T$ and the corresponding camera pose $\Xi \in \mathrm{SE(3)}$ obtained during initialization (see Sec. 3.2 in the main paper).

Using the estimated virtual depths $v_i$ and calculated image distances $b_{Li}$ ($i \in  \{1,\dotsc,N\}$) for $N$ points, the following linear system of equations can be defined (\cref{eq:systemBbL0}). 

\begin{align}
\begin{bmatrix} 
    b_{L1}\\
    b_{L2}\\
    \vdots\\
    b_{LN}
    \end{bmatrix} &= 
\begin{bmatrix} 
    v_{1} & 1\\
    v_{2} & 1\\
    \vdots & \vdots\\
    v_{N} & 1
    \end{bmatrix}
    \cdot \begin{bmatrix}
B\\
b_{L0}
\end{bmatrix}\nonumber\\
    B_L &= V \cdot \begin{bmatrix}
B\\
b_{L0}
\end{bmatrix}
  \label{eq:systemBbL0}
\end{align}
Initialization for the parameters $B$ and $b_{L0}$ is obtained as a standard least-squares solution defined as follows:
\begin{align}
    \begin{bmatrix}
    B\\
    b_{L0}
    \end{bmatrix} &= (V^T \cdot V)^{-1} \cdot V^T \cdot B_L,\nonumber\\
\begin{bmatrix}
B\\
b_{L0}
\end{bmatrix} &= \left(\begin{bmatrix} 
    v_{1} & v_{2} & \cdots & v_{N}\\
    1 & 1 & \cdots & 1
    \end{bmatrix} \cdot \begin{bmatrix} 
    v_{1} & 1\\
    v_{2} & 1\\
    \vdots & \vdots\\
    v_{N} & 1
    \end{bmatrix} \right)^{-1} \cdot
    \begin{bmatrix} 
    v_{1} & v_{2} & \cdots & v_{N}\\
    1 & 1 & \cdots & 1
    \end{bmatrix} \cdot
    \begin{bmatrix} 
    b_{L1}\\
    b_{L2}\\
    \vdots\\
    b_{LN}
    \end{bmatrix}.
  \label{eq:lssolution}
\end{align}

\section{Sequences for calibration experiments}
\label{sec:SsequencesExperiments}

In this section, we show exemplary extracts from the totally focused images of the sequences used in each experiment provided in the evaluation section of the main paper (Sec. 4). \cref{tab:associationExperimentSequences} summarizes the sequences used for each experiment.

\setlength{\tabcolsep}{9pt}
\begin{table}[H]
  \centering
  \caption{Association of image sequence extracts used in each calibration evaluation experiment with our LiFCal method.}
  \label{tab:associationExperimentSequences}
  \begin{tabular}{c | c c c c c}
    \toprule
    Experiment & Sequences used \\
    \midrule
    Calibration based on a 3D calibration target & \cref{fig:imagesCalibrationTarget}\\
    Online calibration on target-free scenes & \cref{fig:imagesCalibrationAnyScene}\\
    Online recalibration & \cref{fig:imagesCalibrationAnyScene}\\
    Generalization to a different camera & \cref{fig:imagesCalibrationMarkers}\\
    \bottomrule
  \end{tabular}
\end{table}

\begin{figure}[H]
  \centering
   \input{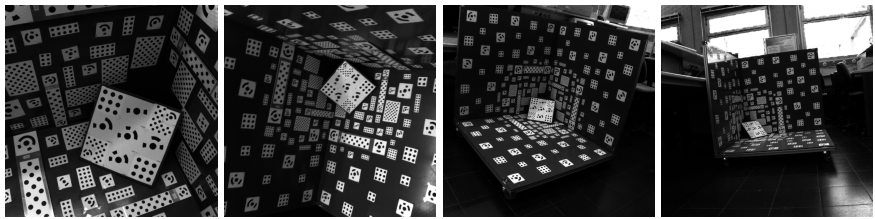}

   \caption{Images of the sequence with the 3D target used in~\cite{zeller2018dissertation}. 94 images for the 12.5~mm lens, 70 images for the 16~mm lens and 76 images for the 35~mm lens.}
   \label{fig:imagesCalibrationTarget}
\end{figure}

\begin{figure}[H]
\centering
\begin{minipage}[b]{.48\textwidth}
    \begin{figure}[H]
      \centering
    
      \adjustbox{minipage=1.3em,valign=t}{\subcaption{}\label{fig:imagesCalibrationAnyScene-a}}%
      \begin{subfigure}[t,valign=t]{\dimexpr1.0\linewidth-1.3em\relax}
      \centering
      \input{images/seq_any_scene/seq_004_lab/seq_004_lab}
      \end{subfigure}%
      \vfill \vspace{2mm}
      \adjustbox{minipage=1.3em,valign=t}{\subcaption{}\label{fig:imagesCalibrationAnyScene-b}}%
      \begin{subfigure}[t,valign=t]{\dimexpr1.0\linewidth-1.3em\relax}
      \centering
      \input{images/seq_any_scene/seq_007_stand_with_papers/seq_007_stand_with_papers}
      \end{subfigure}%
      \vfill \vspace{2mm}
      \adjustbox{minipage=1.3em,valign=t}{\subcaption{}\label{fig:imagesCalibrationAnyScene-c}}%
      \begin{subfigure}[t,valign=t]{\dimexpr1.0\linewidth-1.3em\relax}
      \centering
      \input{images/seq_any_scene/seq_009_phone/seq_009_phone}
      \end{subfigure}%
       \caption{Sample of the totally focused images of the sequence for online calibration: (\protect\subref{fig:imagesCalibrationAnyScene-a}) Lab (56 images from seq\_004), (\protect\subref{fig:imagesCalibrationAnyScene-b}) Hallway (70 images from seq\_007), (\protect\subref{fig:imagesCalibrationAnyScene-c}) Office (76 images from seq\_009).}
       \label{fig:imagesCalibrationAnyScene}
    \end{figure}
\end{minipage}\hfill
\begin{minipage}[b]{.48\textwidth}
    \begin{figure}[H]
      \centering
    
      \adjustbox{minipage=1.3em,valign=t}{\subcaption{}\label{fig:imagesCalibrationMarkers-a}}%
      \begin{subfigure}[t,valign=t]{\dimexpr1.0\linewidth-1.3em\relax}
      \centering
      \input{images/seq_aruco/seq_1_table/seq_1_table.tex}
      \end{subfigure}%
      \vfill \vspace{2mm}
      \adjustbox{minipage=1.3em,valign=t}{\subcaption{}\label{fig:imagesCalibrationMarkers-b}}%
      \begin{subfigure}[t,valign=t]{\dimexpr1.0\linewidth-1.3em\relax}
      \centering
      \input{images/seq_aruco/seq_2_phone/seq_2_phone.tex}
      \end{subfigure}%
      \vfill \vspace{2mm}
      \adjustbox{minipage=1.3em,valign=t}{\subcaption{}\label{fig:imagesCalibrationMarkers-c}}%
      \begin{subfigure}[t,valign=t]{\dimexpr1.0\linewidth-1.3em\relax}
      \centering
      \input{images/seq_aruco/seq_3_computer/seq_3_computer.tex}
      \end{subfigure}%
       \caption{Sample of the totally focused images of the sequence for calibration on any scene using Aruco markers for scaling: (\protect\subref{fig:imagesCalibrationMarkers-a}) Table (42 images), (\protect\subref{fig:imagesCalibrationMarkers-b}) Phone (51 images), (\protect\subref{fig:imagesCalibrationMarkers-c}) Keyboard (50 images).}
       \label{fig:imagesCalibrationMarkers}
    \end{figure}
\end{minipage}
\end{figure}

\section{Reducing the number of features in the scene}
\label{sec:ReducingFeatures}

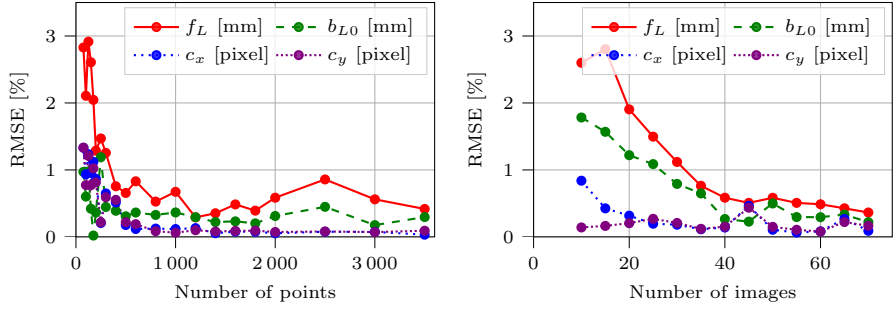
\begin{figure}
\centering
\pgfkeys{/pgf/number format/.cd,1000 sep={\,}}
\begin{tikzpicture}

\tikzstyle{every node}=[font=\scriptsize]

\begin{axis}[
  width=6.35cm,
height=4.7cm,
legend cell align={left},
legend style={
  fill opacity=0.8,
  draw opacity=1,
  text opacity=1,
  anchor=north east,
  draw=white!80!black,
  legend columns=2,
  /tikz/column 2/.style={
                column sep=0pt,
            }
},
tick align=outside,
tick pos=left,
x grid style={white!69.0196078431373!black},
xlabel={Number of points},
xmajorgrids,
xmin=0, xmax=3600,
xtick style={color=black},
y grid style={white!69.0196078431373!black},
ylabel={RMSE [\%]},
ymajorgrids,
ymin=0, ymax=3.5,
ytick style={color=black}
]
\addplot [thick, red, mark=*, mark size=1.5, mark options={solid}]
table {%
75 2.828
100 2.107
125 2.915
150 2.608
175 2.046
200 1.288
250 1.470
300 1.255
400 0.756
500 0.656
600 0.830
800 0.527
1000 0.673
1200 0.292
1400 0.353
1600 0.485
1800 0.392
2000 0.585
2500 0.858
3000 0.561
3500 0.417
};
\addlegendentry{$f_L$ [mm]}
\addplot [thick, green!50!black, mark=*, mark size=1.5, mark options={solid}, dashed]
table {%
75 0.971
100 0.602
125 0.977
150 0.421
175 0.020
200 0.363
250 1.190
300 0.447
400 0.390
500 0.308
600 0.364
800 0.328
1000 0.365
1200 0.294
1400 0.222
1600 0.232
1800 0.202
2000 0.312
2500 0.449
3000 0.176
3500 0.296
};
\addlegendentry{$b_{L0}$ [mm]}
\addplot [thick, blue, mark=*, mark size=1.5, mark options={solid}, dotted]
table {%
75 1.331
100 0.934
125 1.235
150 0.956
175 1.120
200 0.868
250 0.207
300 0.649
400 0.514
500 0.177
600 0.118
800 0.124
1000 0.117
1200 0.131
1400 0.055
1600 0.079
1800 0.077
2000 0.054
2500 0.077
3000 0.074
3500 0.033
};
\addlegendentry{$c_x$ [pixel]}
\addplot [thick, violet, mark=*, mark size=1.5, mark options={solid}, densely dotted]
table {%
75 1.332
100 0.774
125 1.210
150 0.772
175 1.023
200 0.814
250 0.218
300 0.592
400 0.553
500 0.219
600 0.190
800 0.085
1000 0.063
1200 0.100
1400 0.080
1600 0.089
1800 0.093
2000 0.073
2500 0.081
3000 0.072
3500 0.092
};
\addlegendentry{$c_y$ [pixel]}
\end{axis}
\end{tikzpicture}
\pgfkeys{/pgf/number format/.cd,1000 sep={\,}}
\begin{tikzpicture}

\tikzstyle{every node}=[font=\scriptsize]

\begin{axis}[
  width=6.35cm,
height=4.7cm,
legend cell align={left},
legend style={
  fill opacity=0.8,
  draw opacity=1,
  text opacity=1,
  anchor=north east,
  draw=white!80!black,
  legend columns=2,
  /tikz/column 2/.style={
                column sep=0pt,
            }
},
tick align=outside,
tick pos=left,
x grid style={white!69.0196078431373!black},
xlabel={Number of images},
xmajorgrids,
xmin=0, xmax=75,
xtick style={color=black},
y grid style={white!69.0196078431373!black},
ylabel={RMSE [\%]},
ymajorgrids,
ymin=0, ymax=3.5,
ytick style={color=black}
]
\addplot [thick, red, mark=*, mark size=1.5, mark options={solid}]
table {%
10 2.600
15 2.804
20 1.904
25 1.497
30 1.119
35 0.763
40 0.585
45 0.508
50 0.584
55 0.508
60 0.487
65 0.427
70 0.364
};
\addlegendentry{$f_L$ [mm]}
\addplot [thick, green!50!black, mark=*, mark size=1.5, mark options={solid}, dashed]
table {%
10 1.783
15 1.569
20 1.220
25 1.087
30 0.791
35 0.649
40 0.267
45 0.228
50 0.499
55 0.296
60 0.294
65 0.334
70 0.220
};
\addlegendentry{$b_{L0}$ [mm]}
\addplot [thick, blue, mark=*, mark size=1.5, mark options={solid}, dotted]
table {%
10 0.839
15 0.426
20 0.317
25 0.197
30 0.182
35 0.114
40 0.140
45 0.464
50 0.107
55 0.065
60 0.083
65 0.269
70 0.091
};
\addlegendentry{$c_x$ [pixel]}
\addplot [thick, violet, mark=*, mark size=1.5, mark options={solid}, densely dotted]
table {%
10 0.141
15 0.164
20 0.205
25 0.270
30 0.208
35 0.120
40 0.153
45 0.436
50 0.150
55 0.106
60 0.080
65 0.222
70 0.165
};
\addlegendentry{$c_y$ [pixel]}
\end{axis}
\end{tikzpicture}
\caption{\ac{rmse} over 10 runs (in percentage of the ground truth) relative to the number of points (with 30 images) (\textit{left}) and relative to the number of images (with 1500 points) (\textit{right}) used for bundle adjustment with the 12.5~mm lens.}
\label{error_graphs_12mm}
\end{figure}

To demonstrate the robustness of the calibration pipeline even with fewer features, it was applied by varying the number of points and the number of images used. The data used are those with the calibration target (see example images in \cref{fig:imagesCalibrationTarget}), so that the \ac{rmse} can be calculated in relation to the reference given in~\cite{zeller2018dissertation}. The experiment was carried out with three different lenses mounted on the R5 camera: $f_L = 12.5$~mm (\cref{error_graphs_12mm}), $f_L = 16$~mm (\cref{error_graphs_16mm}), and $f_L = 35$~mm (\cref{error_graphs_35mm}).

With sufficient images and points, our LiFCal method achieves an RMSE of around 0.5\% compared with the ground truth. Errors are slightly higher for the 35 mm lens, which can be explained by the inferior scattering of points in the image due to the higher focal length. The data starts to become a little less precise with less than 800 points and fewer than 25 images. Nevertheless, correct calibration data are still obtained down to 75 points and 15 images.

\begin{figure}
\centering
\pgfkeys{/pgf/number format/.cd,1000 sep={\,}}
\begin{tikzpicture}

\tikzstyle{every node}=[font=\scriptsize]

\begin{axis}[
  width=6.35cm,
height=4.7cm,
legend cell align={left},
legend style={
  fill opacity=0.8,
  draw opacity=1,
  text opacity=1,
  anchor=north east,
  draw=white!80!black,
  legend columns=2,
  /tikz/column 2/.style={
                column sep=0pt,
            }
},
tick align=outside,
tick pos=left,
x grid style={white!69.0196078431373!black},
xlabel={Number of points},
xmajorgrids,
xmin=0, xmax=3600,
xtick style={color=black},
y grid style={white!69.0196078431373!black},
ylabel={RMSE [\%]},
ymajorgrids,
ymin=0, ymax=3.5,
ytick style={color=black}
]
\addplot [thick, red, mark=*, mark size=1.5, mark options={solid}]
table {%
50 1.259
75 0.849
100 0.831
125 0.861
150 0.902
175 0.837
200 0.428
250 0.768
300 0.601
400 0.572
500 0.581
600 0.536
800 0.385
1000 0.392
1200 0.242
1400 0.291
1600 0.285
1800 0.171
2000 0.308
2500 0.349
3000 0.235
3500 0.351
};
\addlegendentry{$f_L$ [mm]}
\addplot [thick, green!50!black, mark=*, mark size=1.5, mark options={solid}, dashed]
table {%
50 0.718
75 0.610
100 0.680
125 0.585
150 0.357
175 0.563
200 0.196
250 0.949
300 0.544
400 0.552
500 0.447
600 0.239
800 0.150
1000 0.105
1200 0.151
1400 0.091
1600 0.225
1800 0.337
2000 0.480
2500 0.460
3000 0.474
3500 0.141
};
\addlegendentry{$b_{L0}$ [mm]}
\addplot [thick, blue, mark=*, mark size=1.5, mark options={solid}, dotted]
table {%
50 1.046
75 1.426
100 0.797
125 1.185
150 0.872
175 0.579
200 0.210
250 0.328
300 0.180
400 0.331
500 0.280
600 0.158
800 0.156
1000 0.154
1200 0.110
1400 0.128
1600 0.069
1800 0.113
2000 0.083
2500 0.100
3000 0.117
3500 0.030
};
\addlegendentry{$c_x$ [pixel]}
\addplot [thick, violet, mark=*, mark size=1.5, mark options={solid}, densely dotted]
table {%
50 2.375
75 0.682
100 1.083
125 0.916
150 0.519
175 0.613
200 0.221
250 0.385
300 0.560
400 0.238
500 0.280
600 0.315
800 0.247
1000 0.181
1200 0.193
1400 0.121
1600 0.194
1800 0.102
2000 0.0886
2500 0.089
3000 0.053
3500 0.158
};
\addlegendentry{$c_y$ [pixel]}
\end{axis}
\end{tikzpicture}
\pgfkeys{/pgf/number format/.cd,1000 sep={\,}}
\begin{tikzpicture}

\tikzstyle{every node}=[font=\scriptsize]

\begin{axis}[
  width=6.35cm,
height=4.7cm,
legend cell align={left},
legend style={
  fill opacity=0.8,
  draw opacity=1,
  text opacity=1,
  anchor=north east,
  draw=white!80!black,
  legend columns=2,
  /tikz/column 2/.style={
                column sep=0pt,
            }
},
tick align=outside,
tick pos=left,
x grid style={white!69.0196078431373!black},
xlabel={Number of images},
xmajorgrids,
xmin=0, xmax=75,
xtick style={color=black},
y grid style={white!69.0196078431373!black},
ylabel={RMSE [\%]},
ymajorgrids,
ymin=0, ymax=3.5,
ytick style={color=black}
]
\addplot [thick, red, mark=*, mark size=1.5, mark options={solid}]
table {%
8 2.444
10 1.219
15 1.064
20 0.659
25 0.436
30 0.270
35 0.337
40 0.324
45 0.372
50 0.415
55 0.191
60 0.301
65 0.186
70 0.092
};
\addlegendentry{$f_L$ [mm]}
\addplot [thick, green!50!black, mark=*, mark size=1.5, mark options={solid}, dashed]
table {%
8 1.040
10 0.846
15 0.592
20 0.350
25 0.219
30 0.226
35 0.361
40 0.260
45 0.139
50 0.195
55 0.185
60 0.140
65 0.145
70 0.076
};
\addlegendentry{$b_{L0}$ [mm]}
\addplot [thick, blue, mark=*, mark size=1.5, mark options={solid}, dotted]
table {%
8 0.484
10 0.841
15 0.487
20 0.365
25 0.119
30 0.170
35 0.181
40 0.218
45 0.213
50 0.135
55 0.184
60 0.115
65 0.075
70 0.100
};
\addlegendentry{$c_x$ [pixel]}
\addplot [thick, violet, mark=*, mark size=1.5, mark options={solid}, densely dotted]
table {%
8 1.503
10 0.722
15 0.474
20 0.335
25 0.129
30 0.143
35 0.197
40 0.248
45 0.185
50 0.063
55 0.170
60 0.107
65 0.134
70 0.067
};
\addlegendentry{$c_y$ [pixel]}
\end{axis}
\end{tikzpicture}
\caption{\ac{rmse} over 10 runs (in percentage of the ground truth) relative to the number of points (with 30 images) (\textit{left}) and relative to the number of images (with 1500 points) (\textit{right}) used for bundle adjustment with the 16~mm lens.}
\label{error_graphs_16mm}
\end{figure}

\begin{figure}
\centering
\pgfkeys{/pgf/number format/.cd,1000 sep={\,}}
\begin{tikzpicture}

\tikzstyle{every node}=[font=\scriptsize]

\begin{axis}[
  width=6.35cm,
height=4.7cm,
legend cell align={left},
legend style={
  fill opacity=0.8,
  draw opacity=1,
  text opacity=1,
  anchor=north east,
  draw=white!80!black,
  legend columns=2,
  /tikz/column 2/.style={
                column sep=0pt,
            }
},
tick align=outside,
tick pos=left,
x grid style={white!69.0196078431373!black},
xlabel={Number of points},
xmajorgrids,
xmin=0, xmax=3600,
xtick style={color=black},
y grid style={white!69.0196078431373!black},
ylabel={RMSE [\%]},
ymajorgrids,
ymin=0, ymax=3.5,
ytick style={color=black}
]
\addplot [thick, red, mark=*, mark size=1.5, mark options={solid}]
table {%
50 3.079
75 0.885
100 1.161
125 0.780
150 0.994
175 0.810
200 0.842
250 0.870
300 0.939
400 0.685
500 0.900
600 0.686
800 0.907
1000 0.890
1200 1.159
1400 0.938
1600 0.785
1800 0.760
2000 0.940
2500 0.735
3000 0.896
3500 0.794
};
\addlegendentry{$f_L$ [mm]}
\addplot [thick, green!50!black, mark=*, mark size=1.5, mark options={solid}, dashed]
table {%
50 3.238
75 1.088
100 1.236
125 0.504
150 1.146
175 0.807
200 0.679
250 0.773
300 1.122
400 0.725
500 0.690
600 0.644
800 0.323
1000 0.641
1200 0.553
1400 0.811
1600 0.528
1800 0.820
2000 0.642
2500 0.670
3000 0.670
3500 0.627
};
\addlegendentry{$b_{L0}$ [mm]}
\addlegendentry{$B$ [mm]}
\addplot [thick, blue, mark=*, mark size=1.5, mark options={solid}, dotted]
table {%
50 4.124
75 3.227
100 1.793
125 3.141
150 3.167
175 1.668
200 2.773
250 1.909
300 1.422
400 2.281
500 1.701
600 1.616
800 1.057
1000 0.711
1200 0.963
1400 1.375
1600 1.065
1800 1.013
2000 1.066
2500 0.353
3000 1.014
3500 0.909
};
\addlegendentry{$c_x$ [pixel]}
\addplot [thick, violet, mark=*, mark size=1.5, mark options={solid}, densely dotted]
table {%
50 7.335
75 1.607
100 2.904
125 1.768
150 1.583
175 1.686
200 2.153
250 0.958
300 1.503
400 0.937
500 1.738
600 1.339
800 1.054
1000 1.094
1200 0.977
1400 0.711
1600 1.065
1800 0.627
2000 0.791
2500 0.648
3000 0.485
3500 0.358
};
\addlegendentry{$c_y$ [pixel]}
\end{axis}
\end{tikzpicture}
\pgfkeys{/pgf/number format/.cd,1000 sep={\,}}
\begin{tikzpicture}

\tikzstyle{every node}=[font=\scriptsize]

\begin{axis}[
  width=6.35cm,
height=4.7cm,
legend cell align={left},
legend style={
  fill opacity=0.8,
  draw opacity=1,
  text opacity=1,
  anchor=north east,
  draw=white!80!black,
  legend columns=2,
  /tikz/column 2/.style={
                column sep=0pt,
            }
},
tick align=outside,
tick pos=left,
x grid style={white!69.0196078431373!black},
xlabel={Number of images},
xmajorgrids,
xmin=0, xmax=75,
xtick style={color=black},
y grid style={white!69.0196078431373!black},
ylabel={RMSE [\%]},
ymajorgrids,
ymin=0, ymax=3.5,
ytick style={color=black}
]
\addplot [thick, red, mark=*, mark size=1.5, mark options={solid}]
table {%
10 6.025
15 2.985
20 2.490
25 0.908
30 1.415
35 1.047
40 1.108
45 0.778
50 0.379
55 0.458
60 0.662
65 0.610
70 0.609
};
\addlegendentry{$f_L$ [mm]}
\addplot [thick, green!50!black, mark=*, mark size=1.5, mark options={solid}, dashed]
table {%
10 3.879
15 1.231
20 0.686
25 0.534
30 0.413
35 0.682
40 0.583
45 0.851
50 0.670
55 0.347
60 0.937
65 0.490
70 0.566
};
\addlegendentry{$b_{L0}$ [mm]}
\addplot [thick, blue, mark=*, mark size=1.5, mark options={solid}, dotted]
table {%
10 9.150
15 5.397
20 0.710
25 1.495
30 0.741
35 0.849
40 1.181
45 0.821
50 0.401
55 0.529
60 0.503
65 0.477
70 0.436
};
\addlegendentry{$c_x$ [pixel]}
\addplot [thick, violet, mark=*, mark size=1.5, mark options={solid}, densely dotted]
table {%
10 19.094
15 1.496
20 0.890
25 1.029
30 1.069
35 0.874
40 0.840
45 0.657
50 0.966
55 0.471
60 0.619
65 0.328
70 0.365
};
\addlegendentry{$c_y$ [pixel]}
\end{axis}
\end{tikzpicture}
\caption{\ac{rmse} over 10 runs (in percentage of the ground truth) relative to the number of points (with 30 images) (\textit{left}) and relative to the number of images (with 1500 points) (\textit{right}) used for bundle adjustment with the 35~mm lens.}
\label{error_graphs_35mm}
\end{figure}

\end{document}